\documentclass[fleqn,usenatbib]{mnras}

\usepackage{newtxtext,newtxmath}
\usepackage{tabularx}
\usepackage{booktabs}
\usepackage{makecell}
\usepackage{booktabs}
\usepackage{threeparttable}   
\usepackage{siunitx}          
\usepackage{threeparttable}
\makeatletter
\@ifundefined{abovecaptionskip}{\newlength\abovecaptionskip\setlength\abovecaptionskip{0pt}}{}
\@ifundefined{belowcaptionskip}{\newlength\belowcaptionskip\setlength\belowcaptionskip{0pt}}{}
\makeatother

\usepackage[T1]{fontenc}

\DeclareRobustCommand{\VAN}[3]{#2}
\let\VANthebibliography\thebibliography
\def\thebibliography{\DeclareRobustCommand{\VAN}[3]{##3}\VANthebibliography}

\usepackage{graphicx}	
\usepackage{amsmath}	
\usepackage{orcidlink}

\title[Tiny Galaxies, Mighty Nitrogen]{Tiny Galaxies, Mighty Nitrogen: Linking Chemical Anomalies to Compact Galaxies at High Redshift using JWST/JADES}

\author[Thakurdesai et al.]{Urvi Thakurdesai\orcidlink{0009-0008-9556-3821},$^{1,2}$\thanks{E-mail: umt20@cam.ac.uk (UT)} 
Xihan Ji\orcidlink{0000-0002-1660-9502},$^{2,3}$
Yuki Isobe\orcidlink{0000-0001-7730-8634},$^{2,3}$
Roberto Maiolino\orcidlink{0000-0002-4985-3819},$^{2,3,4}$
Vasily Belokurov\orcidlink{0000-0002-0038-9584},$^{1}$
\and
Courtney Carreira\orcidlink{0000-0001-6301-3667},$^{5}$ Brant Robertson\orcidlink{0000-0002-4271-0364},$^{5}$ 
Will McClymont\orcidlink{0009-0009-5565-3790},$^{2,3}$ D\'avid Pusk\'as\orcidlink{0000-0001-8630-2031},$^{2,3}$ 
Andrew Bunker\orcidlink{0000-0002-8651-9879},$^{6}$ 
\and 
Stephane Charlot\orcidlink{0000-0003-3458-2275},$^{7}$
Mirko Curti\orcidlink{0000-0002-2678-2560},$^{8}$ 
Qiao Duan\orcidlink{0009-0009-8105-4564},$^{2,3}$ 
Francesco D'Eugenio\orcidlink{0000-0003-2388-8172},$^{2,3}$
Maria Koller\orcidlink{0009-0000-1950-9112},$^{2,3}$
\and
Jonah Powley\orcidlink{0009-0008-4758-8071},$^{2,3}$
Pierluigi Rinaldi\orcidlink{0000-0002-5104-8245},$^{9}$
Jan Scholtz\orcidlink{0000-0001-6010-6809},$^{2,3}$ Charlotte Simmonds\orcidlink{0000-0003-4770-7516},$^{10}$ 
Sandro Tacchella\orcidlink{0000-0002-8224-4505},$^{2,3}$ and
\and
Hannah $\ddot{\rm U}$bler\orcidlink{0000-0003-4891-0794}$^{11}$ 
\\
\\
$^{1}$Institute of Astronomy, University of Cambridge, Madingley Road, Cambridge CB3 0HA, UK\\
$^{2}$Kavli Institute for Cosmology, University of Cambridge, Madingley Road, Cambridge, CB3 0HA, UK\\
$^{3}$Cavendish Laboratory, University of Cambridge, 19 JJ Thomson Avenue, Cambridge CB3 0HE, UK\\
$^{4}$Department of Physics and Astronomy, University College London, Gower Street, London WC1E 6BT, UK\\
$^{5}$Department of Astronomy and Astrophysics, University of California, Santa Cruz, 1156 High Street, Santa Cruz, CA 95064, USA\\
$^{6}$Department of Physics, University of Oxford, Denys Wilkinson Building, Keble Road, Oxford OX1 3RH, UK\\
$^{7}$Sorbonne Universit\'e, CNRS, UMR 7095, Institut d'Astrophysique de Paris, 98 bis bd Arago, 75014 Paris, France\\
$^{8}$Department of Astronomy \& Cosmic Frontier Center, The University of Texas at Austin, Austin, TX 78712, USA\\
$^{9}$Space Telescope Science Institute, 3700 San Martin Drive, Baltimore, Maryland 21218, USA\\
$^{10}$Departamento de Astronomía, Universidad de Chile, Camino El Observatorio 1515, Las Condes, 7591245 Santiago, Chile\\
$^{11}$Max-Planck-Institut f$\ddot{ u}$r extraterrestrische Physik, Gießenbachstraße 1, Garching, 85748, Germany
}

\date{Accepted XXX. Received YYY; in original form ZZZ}

\pubyear{\the\year{}}

\begin{document}
\label{firstpage}
\pagerange{\pageref{firstpage}--\pageref{lastpage}}
\maketitle

\begin{abstract}
The origin of anomalous chemical enrichment in high-redshift galaxies discovered by JWST remains a key question in galaxy evolution. Enhanced nitrogen abundances have been identified in several galaxies at early cosmic times, but the physical conditions driving this enrichment, and their connection to galaxy morphology, remain unclear. We test whether morphologically compact galaxies at $4<z<7$ show stronger chemical anomalies than extended systems. Using JWST/NIRSpec spectroscopic data, we construct samples of compact and extended galaxies and analyse their rest-frame spectra. We stack the spectra of each population and measure key UV and optical emission lines. We detect nitrogen emission in the compact galaxy stack, including N~\textsc{iv}], and [N~\textsc{ii}], while also detecting key carbon and oxygen lines. The detection of [O~\textsc{iii}]$\lambda4363$ enables a direct-$T_{\rm e}$ determination of the oxygen abundance. From these measurements, we infer chemical abundance ratios, including $\log({\rm N/O})$ and $\log({\rm C/O})$. We find that compact galaxies exhibit elevated $\log({\rm N/O})$ (by $\sim0.7$ dex) and $\log({\rm C/O})$ (by $\sim0.5$ dex) abundance ratios relative to extended systems. The abundance patterns of compact galaxy stacks are comparable to those of previously identified individual nitrogen-enhanced galaxies at high redshift and second-generation globular-cluster stars in the Milky Way. These results suggest a possible link between galaxy compactness and chemical enrichment pathways in the early Universe. Deeper JWST NIRSpec IFU observations of compact high-redshift galaxies will be crucial for determining the origin, spatial distribution, and evolutionary significance of these abundance patterns.
\end{abstract}




\section{Introduction}\label{sec:intro}

The early Universe consisted of pristine gas, composed of hydrogen, helium, and traces of lithium, produced during the Big Bang nucleosynthesis \citep{hoyle_mystery_1964, peebles_primordial_1966, wagoner_synthesis_1967, cyburt_big_2016}. Small primordial density perturbations grew under gravity, leading to the formation of dark matter halos that provided the potential wells in which gas could collapse and the first stars and galaxies could form \citep{barkana_scale-dependent_2011, bromm_first_2011, klessen_first_2023}. As this gas cooled and condensed within these halos, the first generation of stars and galaxies were formed \citep{klessen_first_2023}. Stellar nucleosynthesis subsequently produced metals, elements heavier than helium, enriching the interstellar medium (ISM) and thus driving the chemical evolution of galaxies over cosmic time. 

The production and dispersal of metals provide fundamental insights into galaxy evolution by tracing star formation, stellar feedback, and black hole growth in the galaxy \citep[see review by,][]{maiolino_re_2019}. The metal content of galaxies, commonly referred to as metallicity, quantifies the abundance of elements heavier than helium relative to hydrogen. In this work, we focus on the gas-phase oxygen abundance, $12+\log(\mathrm{O/H})$, which we use as an observational proxy for gas-phase metallicity. Oxygen is particularly useful as it is one of the most abundant metals and its strong emission lines provide accessible and widely used diagnostics of the chemical state of ionised gas. The metals are synthesised in the stars and are expelled in the ISM through stellar winds and supernovae explosions, especially core collapse supernovae (CCSNe), which drive the metals out further. This enriches the surrounding gas with newly synthesized heavy elements \citep{izotov_heavy_1999, berg_chemical_2019, kobayashi_origin_2020, kobayashi_chemo-dynamical_2023}.

Consequently, metallicity is regulated by the interplay between star formation and inflow of pristine gas which can dilute the chemical content of the ISM \citep{kobayashi_origin_2020, koeppen_evolution_1994, keres_how_2005}. Hence, these chemical abundances preserve a record of their past evolution \citep{dave_galaxy_2011, lilly_gas_2013}. Understanding the evolution of metallicity is therefore essential for constraining the processes that drive galaxy formation and evolution. Further studying the chemical abundances on spatially resolved scales provides a powerful means of probing the assembly of stellar populations and the distribution of metals across the ISM \citep{curti_massmetallicity_2020,maiolino_re_2019}.

Among the known scaling relations in chemical abundances, one of the most widely studied is that between the gas phase oxygen abundance (O/H) and nitrogen-to-oxygen abundance ratio (hereafter N/O). Since oxygen and nitrogen are synthesized by stars on different timescales, the N/O ratio traces the yields of the dying stars across various masses, which helps us study the star formation histories of galaxies \citep{izotov_chemical_2006, vincenzo_nitrogen_2016}. Previous such studies have shown that N/O generally increases with gas-phase metallicity (O/H). This correlation has been observed between the O/H and N/O in the local H~\textsc{ii} regions and in the ISM of star-forming (SF) galaxies \citep{vila-costas_nitrogen--oxygen_1993, zee_spectroscopy_1998, pilyugin_abundance_2012, andrews_mass-metallicity_2013, berg_carbon_2016, berg_chemical_2019, berg_chaos_2020, belfiore_sdss_2017, nicholls_abundance_2016, hayden-pawson_klever_2022}.

In local galaxies, the nitrogen-to-oxygen abundance ratio exhibits little or no dependence on gas-phase oxygen abundance at low metallicities, \(12+\log(\mathrm{O/H})\lesssim8.0\), producing the well-known low-metallicity N/O plateau. This behaviour is generally attributed to the contribution of primary nitrogen, with massive stars providing a prompt source of primary N \citep{izotov_heavy_1999, perez-montero_impact_2009, berg_chemical_2019, arellano-cordova_jwst_2025, scholte_electron_2026}. Towards higher oxygen abundances, N/O increases with O/H as the contribution from low- and intermediate-mass stars, which can produce both primary and secondary nitrogen, becomes increasingly important. At the highest oxygen abundances, the N/O--O/H relation can become particularly steep. Chemical-evolution models have proposed several mechanisms for this behaviour, including differential metal loss through galactic winds, in which oxygen may be preferentially removed relative to nitrogen, although the relative importance of such processes remains model-dependent \citep{vila-costas_nitrogen--oxygen_1993, pilyugin_abundance_2012, vincenzo_nitrogen_2016, berg_chaos_2020}.

The change in the behavior of N/O with oxygen abundance is generally interpreted in terms of the different nucleosynthetic origins and metallicity dependencies of nitrogen production. Nitrogen is produced through primary and secondary channels, distinguished by the origin of the C and O nuclei that serve as seeds for the CNO cycle. In primary production, these nuclei are synthesized within the star itself, making the nitrogen yield relatively insensitive to the initial metallicity. In secondary production, the C and O nuclei are inherited from the material from which the star formed, causing the nitrogen yield to increase with initial metallicity \citep[e.g.][]{meynet_stellar_2002, vincenzo_nitrogen_2016, maiolino_re_2019}. Both massive and low- to intermediate-mass stars can contribute to both channels. Massive stars can produce primary N through rotationally induced mixing, particularly at low metallicity, while AGB stars can produce primary N through dredge-up and hot-bottom burning as well as secondary N from pre-existing C and O. Massive stars provide relatively prompt nitrogen enrichment through stellar winds and core-collapse supernovae, whereas low- and intermediate-mass stars contribute on longer timescales through their post-main-sequence evolution. Consequently, elevated N/O ratios are commonly observed in metal-rich local star-forming galaxies, while super-solar N/O at low metallicity is relatively uncommon, with notable exceptions such as Mrk 996 and NGC 5253 \citep[cf.][]{telles_geminigmos_2014, james_vlt_2009, walsh_optical_1989}.


The advent of \textit{James Webb Space Telescope} \citep[JWST, ][]{gardner_james_2023, rigby_science_2023}, has enabled unprecedented sensitivity in the near-infrared with the \textit{Near Infrared Spectrograph} \citep[NIRSpec,][]{jakobsen_near-infrared_2022, boker_-orbit_2023} revolutionizing our understanding of galaxy formation and chemical evolution across cosmic time \citep{stark_observations_2025}. Using the various observing modes of NIRSpec, elemental abundances have been estimated for galaxies upto redshift $z\sim14$ \citep{bunker_jades_2023, cameron_nitrogen_2023, castellano_jwst_2024, napolitano_seven_2025, carniani_eventful_2025, schouws_detection_2025, curti_jades_2025, naidu_cosmic_2025}. One of the most remarkable discoveries from \textit{JWST} has been the detection of extreme nitrogen enhancement in chemically young, metal-poor galaxies, which has challenged standard chemical evolution models. The first evidence came from the detection of strong N~\textsc{iv}]$\lambda\lambda1483,1486$ and N~\textsc{iii}]$\lambda\lambda1747$-$1754$ emission lines in the high redshift galaxy at $z=10.6$, GNz11, indicating a super solar nitrogen abundance of N/O $>-0.25$ at sub solar metallicity of $12 +\rm log(O/H) = 7.91\pm0.07$ \citep{bunker_jades_2023, cameron_nitrogen_2023}.

Subsequent \textit{JWST} observations have identified more than 60
such galaxies at $z>5$ exhibiting super solar nitrogen abundances. Some notable examples are GHZ2 at $z = 12.34$ \citep{castellano_jwst_2024}, GHZ9 at $z =10.15$ \citep{napolitano_seven_2025}, CEERS\_1019 at $z =8.68$ \citep{zamora_ga-nifs_2025, isobe_jwst_2023, marques-chaves_extreme_2024}, MoM-z14 at $z =14.44$ \citep{naidu_cosmic_2025}, and several others \citep{curti_jades_2025, naidu_black_2025, topping_metal-poor_2024, topping_deep_2024, zhang_potential_2025, marques-chaves_extreme_2024, castellano_jwst_2024, labbe_unambiguous_2024, topping_aurora_2025, schaerer_discovery_2024, navarro-carrera_constraints_2024, stiavelli_what_2024, yanagisawa_strong_2024, napolitano_seven_2025, arellano-cordova_classy_2024, ji_connecting_2025, isobe_empress_2023}. Some systems exhibit extremely high nitrogen abundances, reaching $\rm \log(N/O)=0.4\pm0.1$ \citep[GS\_3073;][]{ji_ga-nifs_2024, ubler_ga-nifs_2023}. These N/O-enhanced galaxies (hereinafter, NOEGs; \citealt{ji_connecting_2025}) have chemical abundances which coincide with the abundance anomalies observed in local globular clusters (GCs, \citealp{belokurov_nitrogen_2023, ji_connecting_2025, gratton_what_2019,  bastian_multiple_2018, charbonnel_n-enhancement_2023, pascale_nitrogen-enriched_2023}). The population now extends to lower redshifts, with N/O-enhanced systems also identified at $z\sim4$ \citep{rusakov_diverse_2026, cameron_jades_2026, morel_discovery_2025}. While many of these discoveries have relied on rest-frame UV nitrogen diagnostics, recent observations have also demonstrated that N/O enhancement can be probed through low-ionisation optical emission lines, including [N~\textsc{ii}] and [O~\textsc{ii}], at high redshift \citep{arellano-cordova_classy_2025, stiavelli_what_2025, zhang_potential_2025, cataldi_marta_2025, scholte_electron_2026}. Together, these observations establish N/O enhancement as a phenomenon extending across a broad redshift range and accessible through complementary UV and optical diagnostics.

The discovery of super solar N/O ratios in metal-poor galaxies presents a significant challenge to conventional chemical evolution models. Hence, many studies have proposed different enrichment scenarios. One possibility is localized nitrogen enrichment by Wolf-Rayet (WR) stars, whose stellar winds can temporarily elevate the N/O ratio especially after a starburst in galaxies \citep{kobayashi_rapid_2024, bekki_model_2023, watanabe_empress_2024, berg_fleeting_2025}. Subsequent feedback from CCSNe may drive outflows that remove enriched gas from the galaxy, while the accretion of relatively pristine gas could replenish the ISM and dilute the observed nitrogen abundance. Other studies have suggested that delayed enrichment AGB stars coupled with bursty star formation \citep{rai_unveiling_2026} or very massive stars (VMS, $100-500 M_{\odot}$) could efficiently produce large nitrogen yields \citep{vink_very_2023, higgins_impact_2025}. Alternatively, super-massive stars (SMS, $>1000 M_{\odot}$), which are predicted to form in dense environments, have also been proposed as potential sources of nitrogen enrichment \citep{cameron_nitrogen_2023, charbonnel_n-enhancement_2023, nagele_multiple_2023, ji_ga-nifs_2024, nandal_100010000_2025}. Additional mechanisms include possible differential galactic winds which preferentially eject CCSNe enriched oxygen-rich gas \citep{rizzuti_high_2025}, tidal disruption events \citep[TDEs, ][]{cameron_nitrogen_2023}, and pollution from Population III stars \citep{nandal_100010000_2025}. Finally, cosmological simulations from the \texttt{THESAN-ZOOM} project \citep{kannan_introducing_2025} suggest that nitrogen-rich star-forming regions can form when previously ejected, oxygen-rich gas from CCSNe mixes with nitrogen-enriched gas produced by earlier stellar populations and subsequently collapses into new molecular clouds \citep{mcclymont_span_2026}.

Beyond the specific enrichment mechanisms, the physical environments in which these systems form provide an important clue to the origin of their unusual abundances. In particular, an emerging connection between compact galaxy morphologies and enhanced N/O ratios suggests that nitrogen enrichment may be associated with regions of exceptionally dense and intense star formation \citep{ji_connecting_2025, naidu_cosmic_2025, nandal_100010000_2025}. The known NOEGs in the literature are likewise also inferred to be compact, with characteristic sizes below $200 \ pc$, making them more compact than typical star-forming galaxies, although these size constraints can be uncertain \citep{naidu_cosmic_2025, harikane_jwst_2025, schaerer_discovery_2024, ji_connecting_2025, roberts-borsani_jwst_2026}. Their high star formation rate densities ($\Sigma_{\rm SFR}$) further indicate that star formation is concentrated into small physical regions \citep{james_vlt_2009, pascale_nitrogen-enriched_2023, ji_ga-nifs_2024}. This raises the possibility that the observed N/O enhancement is spatially localized to compact, high gas density star forming regions rather than being representative of the galaxy as a whole. The dense environments inferred for these systems may also resemble those associated with the formation of local globular clusters. However, this comparison is between nebular gas abundances in unresolved high-redshift NOEGs and stellar abundances in surviving local GC stars, so the observed similarities do not by themselves establish a direct evolutionary connection, but may instead point to a possible common enrichment environment between high-redshift NOEGs and the progenitors of present-day GCs \citep{ji_connecting_2025}.

High-redshift AGNs provide a particularly interesting extension of this picture. Type-1 AGNs have frequently been found to exhibit enhanced nitrogen abundances, while many \textit{JWST}-identified AGNs also show compact morphologies \citep{isobe_jades_2025, matthee_little_2024, maiolino_jades_2024, juodzbalis_jades_2026}. However, the nitrogen enhancement measured in broad-line AGN spectra traces dense gas in the nuclear broad-line region and should not be directly equated with the galaxy-wide N/O abundance of the star-forming ISM. Nevertheless, the coexistence of compactness, nitrogen enhancement, and AGN activity may therefore point towards a broader connection between dense stellar environments, proto-globular cluster formation, and the conditions associated with black-hole seeding in the early Universe \citep{isobe_jades_2025}.

Motivated by these observations, we investigate whether relative galaxy compactness is associated with differences in nitrogen abundance at $4<z<7$. To analyse this argument, we use the JWST Advanced Deep Extragalactic Survey (JADES) spectroscopic observations from the Data Release 4 and imaging observations from Data Release 5, to define a parent sample of galaxies in the redshift $4<z<7$ range. We define morphologically compact and extended populations relative to the stellar-normalized galaxy size distribution within redshift bins, allowing us to test the role of compactness at fixed stellar mass and cosmic epoch. We then construct stacked spectra for each population, model their UV and optical nitrogen emission lines, and derive elemental abundances to investigate whether relatively compact galaxies exhibit systematically enhanced nitrogen abundances compared with their more extended counterparts.

We outline the paper as follows. We describe our data and our selection criteria for the morphologically distinct galaxies in Section \ref{sec:data}. In Section \ref{sec:methods}, we present our stacking method and emission line fitting pipeline. In Section \ref{sec:neb_prop}, we describe the derivation of the nebular properties of the stacked spectra and our methods for calculating the elemental abundances. We present and discuss our results in Section \ref{sec:results_disc} and summarize our conclusions in Section \ref{sec:conclusions}. Throughout this work, we assume a flat lambda CDM cosmology with $\rm H_0 = 67.4 \ km/s/Mpc$ and $\Omega_m$ = 0.315 \citep{planck_collaboration_planck_2020}. We adopt the solar abundance, $12 + \rm log(C/H)_{\odot}  = 8.43$, $12 + \rm log(N/H)_{\odot} = 7.83$, and $12 + \rm log(O/H)_{\odot} = 8.69$ \citep{asplund_chemical_2021}. We refer to the relative oxygen abundance, O/H, as the metallicity, unless otherwise specified.

\begin{figure}
	\includegraphics[width=1.0
    \columnwidth]{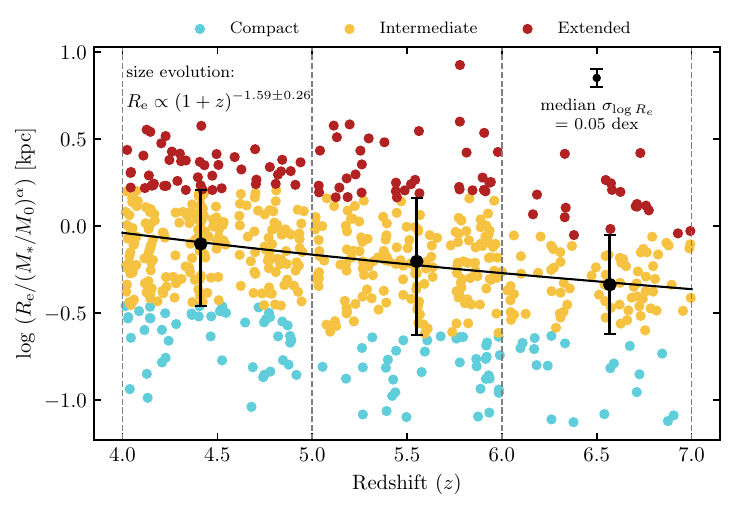}
    \caption{Redshift evolution of the mass-normalized rest-frame UV $R_e$ for the 657 galaxies in our parent JADES DR4 sample. Galaxy sizes are obtained from the JADES DR5 morphology catalogue, while the stellar masses are taken from the JADES DR4 galaxy stellar population catalogue. Representative errors on the galaxy sizes (0.05 dex) are shown on the top right. The black solid line shows the best-fitting size-redshift evolution, parametrized as $\propto (1+z)^{\alpha_{z}}$, with a best-fitting slope of $\alpha_z=-1.59\pm0.26$. The black circles indicate the median mass-normalized sizes measured for each redshift bin which are used to derive the best-fitting relation. The blue-points denote the galaxies with mass-normalized sizes below the 16th percentile in their respective redshift bins and define the \emph{compact} population, while the red points denote galaxies above the 84th percentile and define the \emph{extended} population. Galaxies with sizes in between these two thresholds constitute the \emph{intermediate} population which are shown as yellow points.}
    \label{fig:size_evolution}
\end{figure}

\section{Observational Data}\label{sec:data}

In this section, we outline the observational data used in this work and the criteria adopted to define the morphology-based galaxy populations used throughout our analysis. We analyse the data products obtained from the \textit{JWST} Advanced Deep Extragalactic Survey \citep[JADES,][]{rieke_jwst_2019, eisenstein_jades_2025}, obtained with the JWST/NIRSpec micro shutter array \citep[MSA,][]{jakobsen_near-infrared_2022, ferruit_near-infrared_2022}. Specifically, we use the value-added catalogues produced by the JADES collaboration for the legacy extragalactic GOODS-South (GOODS-S) and GOODS-North (GOODS-N) survey fields comprising observations from PIDs 1180, 1181, 1210, 1286, 1287, 3215, and 4540 \citep{bunker_jades_2024, deugenio_jades_2024, eisenstein_overview_2025, eisenstein_jades_2025}.




Our spectroscopic analysis is based on the complete NIRSpec spectroscopic sample from the JADES Data Release 4 \citep[hereafter DR4;][]{curtis-lake_jades_2026, scholtz_jades_2025} which provides reduced spectra, spectroscopic redshifts and emission-line measurements. We crossmatch this sample with the complete photometric catalogue from the JADES Data Release 5 \citep[hereafter DR5;][]{alberts_jwst_2026, hainline_jwst_2026, johnson_jwst_2026, robertson_jwst_2026, carreira_jwst_2026, duan_jwst_2026}, from which we obtain structural and morphological measurements used to classify galaxies into the different morphology populations analysed in this work.

Since the JADES overview papers and the DR4 and DR5 data release papers describe the survey design, observations, data reduction procedures, and catalogue constructions in details, we provide a summary of the aspects of the data products and measurements that are most relevant to our analysis.

\subsection{JADES DR4 - Spectroscopic sample}

The JADES DR4 spectroscopic catalogue contains 5190 targets observed in the GOODS-S and GOODS-N fields as a part of the following JADES programs: PIDs 1180, 1181, 1210, 1286, 1287, 3215. All targets were observed with the low-resolution NIRSpec prism configuration ($R\sim100$, hereafter R100), while the majority were also observed with the three medium-resolution gratings F070LP/G140M, F170LP/G235M, and F290LP/G395M ($R\sim1000$, hereafter R1000). Approximately 700 sources were allocated long exposure times of $\gtrsim 20$ hours, whereas the remaining $\sim 4400$ sources were observed with exposure times of $\sim 1-3$ hours per grating setting \citep{curtis-lake_jades_2026}. The R100 prism observations generally received integration times that were longer than the corresponding R1000 observations by a factor of 1-4 \citep{deugenio_jades_2024}. 

The spectroscopic data were reduced using the JADES NIRSpec reduction pipeline developed by the European Space Agency (ESA) NIRSpec Science Operations Team \citep{ferruit_near-infrared_2022} together with the NIRSpec Guaranteed Time Observations (GTO) Team \citep{oliveira_preparing_2018}. The reduction pipeline propagates correlated noise throughout the reduction process by incorporating error propagation and variance-conserving resampling  to preserve the statistical properties of the data \citep{dorner_model-based_2016}. 

For each target, DR4 provides one-dimensional spectra extracted using both 3-pixel box-car aperture and the standard 5-pixel full microshutter aperture \citep{deugenio_jades_2024}. Throughout this work, we use the 3-pixel extraction (pixel size = 0.1 arcsec), because it provides higher signal-to-noise ratio (S/N) for compact sources than the 5--pixel extraction while retaining the majority of the source flux. DR4 also includes a flux catalogue, with line fluxes measured using the \textsc{ppxf} software \citep{cappellari_full_2023}, as detailed in \citet{deugenio_jades_2024}. Each redshift has one of five quality flags, where the quality flags `A', `B', and `C' correspond to secure redshifts determined from multiple emission lines and/or the continuum features, while the flags `D' and `E' indicate tentative and/or unconstrained redshifts measurements. 

\subsection{JADES DR5 - Photometric sample}

The JADES DR5 photometric release provides the imaging and source catalogue for the GOODS-S and the GOODS-N fields from the following JADES programs: PIDs 1180, 1181, 1210, 1286, 1287, 4540, complemented by additional JADES public NIRCam imaging dataset described in \citet{johnson_jwst_2026, robertson_jwst_2026}. The release consists of deep, uniform processed mosaics spanning the optical to mid-infrared wavelengths and additionally incorporates \textit{Hubble Space Telescope} (HST) imaging from Hubble Legacy Fields mosaics. The reduction and calibration of the DR5 imaging data is presented in \citet{johnson_jwst_2026} and the construction of the DR5 source catalogue including photometric results is presented in \citet{robertson_jwst_2026}. 

The catalogue provides extensive multi-band photometry from the core JADES NIRCam filters, including the seven wide-band filters F090W, F115W, F150W, F200W, F277W, F356W, and F444W, together with medium-band imaging in F182M, F210M, F335M, F410M, F430M, F460M, and F480M where available. In this work, we use the morphological measurements derived from the F115W imaging, which contains measurements for 278021 targets in the GOODS-S and 146185 targets in the GOODS-N fields.

Galaxy structural and morphological measurements for the JADES DR5 are presented in \citet{carreira_jwst_2026}. They were obtained by implementing MCMC modelling on a single-component Sérsic profile using a Bayesian inference framework \texttt{PySersic} on all sources detected in the JADES photometric results. Inferred morphological properties include measurements such as effective radius, Sérsic index, axis ratio, and positional angle. 

We additionally make use of the DR5 galaxy stellar population catalogue presented in \citet{duan_jwst_2026}. The stellar population parameters were inferred using Bayesian forward modelling code \texttt{Prospector} \citep{johnson_stellar_2021}, accelerated through the neural network emulator \texttt{Parrot} \citep{mathews_as_2023} for rapid evaluation of \texttt{FSPS} stellar population synthesis models \citep{conroy_propagation_2009, conroy_propagation_2010-1, conroy_propagation_2010}. The catalogue provides a wide range of galaxy properties, including redshifts, stellar masses, star formation rates (SFRs), star formation histories (SFHs), stellar ages, dust attenuation and emission, stellar and gas-phase metallicities, and potential AGN contributions. Furthermore, this catalogue implements a novel prior on the SFH, known as the star-forming main-sequence (SFMS) prior which assumes that galaxies form and evolve along the SFMS, with substantial scatter around this relation.

\begin{figure}
	\includegraphics[width=1.0\columnwidth]{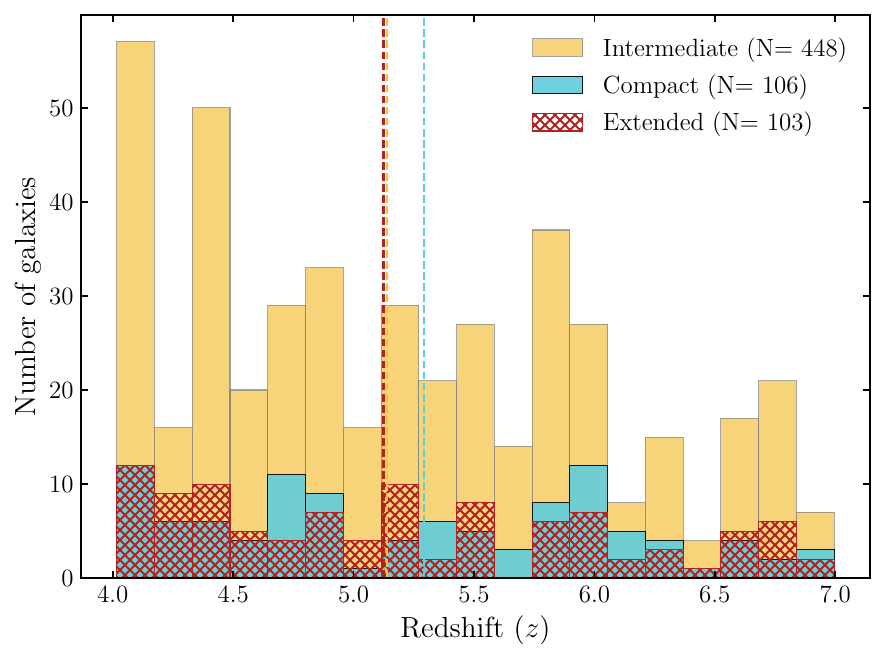}
    \caption{Redshift Distribution of our JADES sample in $4<z<7$ which have reliable redshifts and both R100 and R1000 observations. The blue, red and yellow redshift histograms correspond to the compact, extended and intermediate populations respectively. The dashed lines for each population mark the median redshift of the population.}
    \label{fig:redshift_dist}
\end{figure}

\subsection{Our sample}

\begin{figure*}

	\includegraphics[width=1.0\linewidth]{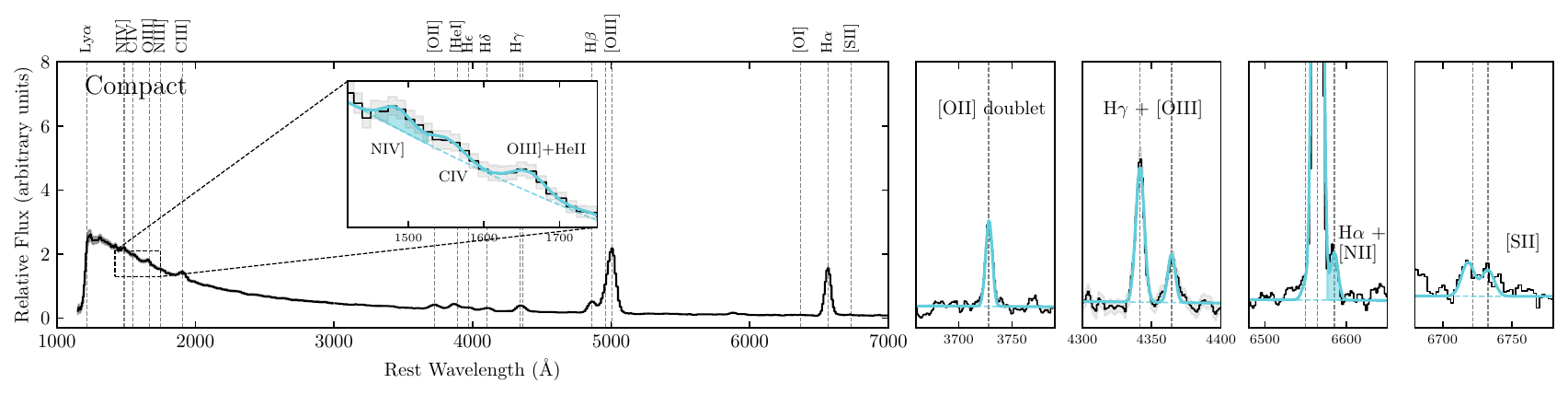}

    \includegraphics[width=1.0\linewidth]{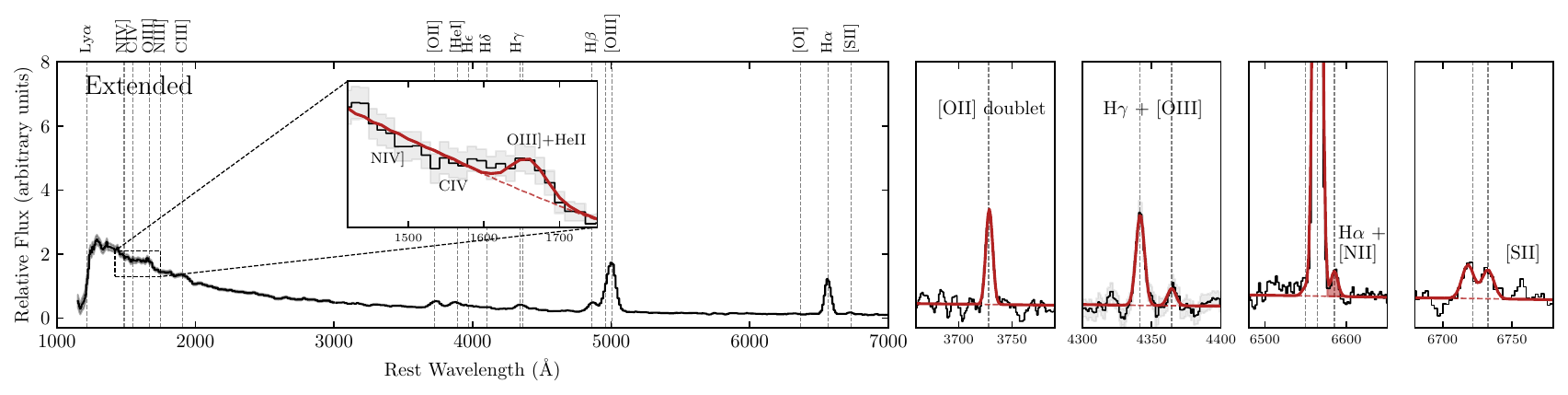}
    \caption{\textbf{Left:} R100 spectra of the median stacking of compact galaxy population (top), and the extended galaxy population (bottom). The inset panels show fitting results of the UV emission lines (blue/red solid curves). \textbf{Right:} R1000 median stacked spectra with pPXF fit for [O~\textsc{ii}]$\lambda\lambda3726,3729$, $\rm H \gamma$ + [O~\textsc{iii}]$\lambda\lambda4363$, $\rm H \alpha + $[N~\textsc{ii}]$\lambda\lambda6548,6584$, and [S~\textsc{ii}]$\lambda\lambda6716,6731$. The shaded regions in both panels highlight the UV and optical nitrogen lines respectively.}
    \label{fig:median_stack}
\end{figure*}

We begin by extracting the one-dimensional spectra from the JADES DR4 spectroscopic data products. To maximize the S/N in the blue regions of the spectra, where the \textit{JWST} point-spread function (PSF) is the most compact, we use the 3-pixel box aperture extractions. This yields us our full spectroscopic sample which consists of 5190 reduced observations. We crossmatch this sample with the JADES DR5 best-fit imaging and morphology catalogue in the F115W filter, which provides the median values of the posterior distributions for each morphological parameter \citep{carreira_jwst_2026}. we further crossmatch these sources with the JADES DR5 galaxy stellar population catalogue using Equatorial coordinates (RA, Dec) \citep{duan_jwst_2026}. This gives us a sample of 4990 spectroscopic sources.

We next apply a series of quality cuts to ensure that the sample consists of galaxies with reliable spectroscopic redshifts ($z$). Specifically, we retain only sources with redshift quality flags 'A', 'B' or 'C', corresponding to secure redshift determinations based on multiple emission lines \citep{deugenio_jades_2024}. We further restrict our sample to the redshift range $4<z<7$, to ensure that both the rest-frame far-ultraviolet (FUV) and rest-frame optical wavelength regions of the galaxy spectrum, including Ly$\alpha$ and H$\alpha$, are covered in both the R100 and R1000 spectra. These selection criteria define our parent sample of 657 galaxies.

To construct our morphology-based galaxy populations, we first analyse the redshift evolution of the ultraviolet (UV) galaxy sizes over our redshift range of $4<z<7$. Since galaxy sizes are known to depend on both stellar masses and redshift as shown in Equation~\ref{eq:size_mass_relation}, we first remove the stellar mass dependence by normalizing the sizes with the stellar masses. 

\begin{equation}\label{eq:size_mass_relation}
    \text{log} R_{\rm e} (M_*,z)/\text{kpc} = \alpha \text{log} (M_*/M_0)+B(z), 
\end{equation}
where we adopt a pivot stellar mass $M_0 = 10^8 M_{\odot}$. The redshift evolution is parametrized as 
\begin{equation}\label{eq:size_mass_relation2}
    B(z) = \beta_z + \alpha_z \text{log}(1+z)
\end{equation}

We present the redshift evolution for the mass-normalized galaxy sizes, $R_{\rm e}/(M_{\ast}/M_{0})^\alpha$, in Figure \ref{fig:size_evolution}. The DR5 morphology catalogue provides the effective (half-light) radius, $R_{\rm e}$, derived from the median of the posterior distribution obtained by fitting a single-component Sérsic profile to each galaxy using \texttt{pysersic} \citep{carreira_jwst_2026}. The reported effective radii correspond to the semi-major axis and are given in arcseconds. We convert these angular sizes into physical units (kpc) using the standard cosmology equations implemented in the \texttt{PYTHON} package \texttt{astropy.cosmology}. The stellar masses, $(M_{\ast}/M_{\odot})$, are obtained from the DR5 galaxy stellar population catalogue, where they are derived at the epoch of the observation using spectral energy distribution (SED) fitting with the Bayesian framework model \texttt{Prospector} \citep{duan_jwst_2026}. 

Using the physical effective radii, $R_e$, in kpc, and the stellar masses, $M_{*}/M_{\odot}$, we calculate the mass-normalized sizes assuming a fixed slope of $\alpha=0.19$, following the best-fitting size-mass relation for star-forming galaxies presented in \citet{morishita_enhanced_2023}. We divide the sample into three redshift bins $4<z<5$, $5<z<6$, and $6<z<7$, and compute the median mass-corrected size in each redshift bin. We then perform a linear regression fit using the \texttt{scipy.stats} package to determine the redshift evolution of the median galaxy sizes.

We obtain a size--redshift evolution of $R_{\rm e}\propto(1+z)^{-1.59\pm0.26}$, indicating a relatively rapid decrease in the average galaxy size with increasing redshift over the range considered in Figure~\ref{fig:size_evolution}. This evolution is steeper than that reported by \citet{morishita_enhanced_2023}, who measured $R_{\rm e}\propto(1+z)^{-0.4\pm0.2}$ for rest-frame UV ($\sim1600$~\AA) sizes over $5<z<14$, and by \citet{shibuya_morphologies_2015}, who found $R_{\rm e}\propto(1+z)^{-1.10\pm0.06}$ for rest-frame UV ($\sim2100$~\AA) sizes over $0<z<7$. \citet{danhaive_beyond_2026}, using \texttt{pysersic} sizes on JADES imaging, find no significant size evolution at fixed mass over $z\approx4$--6 ($R_{\rm e}\propto(1+z)^{0.31\pm0.43}$ in the rest-frame optical), albeit with large uncertainties due to their narrow redshift range. Using the same JADES DR5 catalogue, \citet{carreira_jwst_2026} find a shallower rest-frame optical evolution of $R_{\rm e}\propto(1+z)^{-0.635\pm0.013}$ for galaxies at $z>1$, although their fit is not mass-normalized and is dominated by galaxies at $z\lesssim2$.

We define our galaxy size populations using the mass-normalized size distribution within each redshift bin. Specifically, for each redshift bin, we calculate the 16th and 84th percentiles of the distribution, and classify galaxies below the 16th percentile as \emph{compact} and those above the 84th percentile as \emph{extended}. Galaxies lying between these two thresholds constitute the \emph{intermediate} population. The resulting compact and extended samples contain 106 and 103 galaxies, respectively, while the remaining 448 galaxies comprise the intermediate population. These populations are highlighted in Figure~\ref{fig:size_evolution}, and their corresponding redshift distributions are shown in Figure~\ref{fig:redshift_dist}. The compact, extended, and intermediate samples have median effective radii of $R_{\rm e}=0.215 \pm 0.015$, $2.179 \pm 0.093$, and $0.780 \pm 0.021$~kpc, and median stellar mass surface densities of $\log\Sigma_{\ast}=2.810 \pm 0.616$, $0.776 \pm 0.557$, and $1.877 \pm 0.560$ $M_{\odot}\,{\rm kpc^{-2}}$, respectively, where $\Sigma_{\ast}=M_{\ast}/(2\pi R_{\rm e}^{2})$.


To assess whether the compact population is disproportionately affected by Type-1 AGN emission, which could contribute to the observed nitrogen enhancement, we crossmatch both samples with the Type-1 and Type-2 AGN catalogues of JADES DR4 \citep{juodzbalis_jades_2026, isobe_jades_2025, scholtz_jades_2025}. We identify 36 Type-1 and Type-2 AGNs ($\sim 5\%$) in our parent sample. We identify 13 ($\sim12\%$) in compact, 5 in extended ($\sim5\%$) and 18 AGNs ($\sim3\%$) in intermediate stacks. Because a small AGN fraction does not by itself rule out a significant contribution to a median stack, we repeat our morphological sample selections and stacking after excluding all known Type-1 and Type-2 AGNs. N\textsc{iv}]\,$\lambda1486$ remains detected in the AGN-excluded compact stack at $3\sigma$. The AGN-excluded extended stack, like the full extended stack, shows no  N\textsc{iv}] detection. We therefore conclude that the nitrogen enhancement in compact galaxies is not driven by known Type-1 AGNs.

\section{Methods} \label{sec:methods}

The analysis presented in this paper is based on the compact and extended galaxy populations presented in the sections below. 

\subsection{Spectral Stacking}
\label{sec:stacking}

We begin by constructing stacked spectra of the R100 prism and R1000 medium-resolution grating observations for the compact and extended galaxy populations. Throughout our analysis, we consider our rest-frame wavelength range of $1150-7000$~\AA. All spectra are first shifted to the rest frame using their spectroscopic redshifts, $z$. 

To ensure that all spectra within each population share a common spectral resolution, we degrade each spectrum to the lowest resolution present in that sample. For each spectrum, we compute its native wavelength-dependent resolution from the \textit{nominal} resolution curve of the \textit{JWST}/NIRSpec instrument \citep{jakobsen_near-infrared_2022}, and adopt the broadest line-spread function (LSF) among all spectra in the sample at each wavelength as the target resolution. Each spectrum is then convolved with an additional Gaussian kernel whose width is given by the quadrature difference between the target and native LSF widths, such that the lowest-resolution spectrum remains unchanged while all others are broadened to match it. Since the R1000 is spread across multiple medium grating filters - G140M/F070LP, G235M/F170LP, and G395M/F290LP, we first convolve the spectra within each grating independently using the appropriate wavelength-dependent line-spread function (LSF) \citep{de_graaff_ionised_2024}, implemented with \texttt{gaussian\_filter1d} and then combine the spectra across these filters to produce a continuous R1000 spectrum for each galaxy. 


Following the resolution matching, all spectra in each population are resampled to a common rest-frame wavelength grid using the \texttt{spectres} package \citep{carnall_spectres_2017}, which simultaneously propagates the uncertainties for the resampled spectra from the original individual spectra. The wavelength grid is sampled at half of the full width at half maximum (FWHM) corresponding to the \textit{point-like} NIRSpec spectral resolution, providing Nyquist sampling of the broadened spectra while preserving the common nominal spectral resolution adopted for the stacks. The finer sampling also provides adequate sampling of the emission lines while conserving the spectral resolution of the overall stack. Each spectrum is then continuum-normalized using a relatively line-free window of the spectrum centered at 2100~\AA. We create a narrow box around this pivot wavelength and then fit a local linear continuum. Each spectrum and its uncertainty are divided by the fitted continuum flux density at rest-frame $2100$~\AA.



We construct two types of stacked spectra for each galaxy population. The first is a median stack, obtained by taking the median flux at each wavelength after continuum normalization. The uncertainties on the median stack are estimated using 1000 Monte Carlo realisations. For each of the 1000 trials, every input spectrum is perturbed by drawing random flux values from a Gaussian distribution centered on the measured flux with a standard deviation equal to its corresponding $1\sigma$ uncertainty. The spectra are then re-stacked, and the uncertainty at each wavelength is taken as the standard deviation of the resulting distribution of the stacked fluxes. The Monte Carlo uncertainties reflect only the measurement noise of the individual spectra and do not capture galaxy-to-galaxy variation within each population. We therefore also estimate the uncertainties on the median stack by bootstrap resampling, in which we draw galaxies with replacement from each population and recompute the median stack over 2000 iterations, taking the standard deviation at each wavelength as the uncertainty. The bootstrapped stacks yield results consistent with the Monte Carlo stacks, indicating that our measurements are not driven by a small number of individual galaxies. The second stack is an inverse-variance weighted mean stack, where the flux at each wavelength is computed using the inverse-variance weights. The corresponding uncertainty is calculated from the inverse square root of the summed weights at each wavelength.

Throughout this paper, we primarily present results based on the median stacks with MC error propagations.
The median stacks are less sensitive to a small number of unusually bright galaxies or spectra with exceptionally strong emission lines, making them a robust representation of the typical spectral properties of each galaxy population. The stacked spectra are presented in Figure~\ref{fig:median_stack}.

\begin{table*}
\centering
\begin{threeparttable}

\caption{Emission line fluxes normalized by H$\beta$ flux and S/N measured for the stacked compact galaxies population in median PRISM R100 and medium resolution Grating R1000 spectra.}
\label{tab:linefluxes}

\begin{tabularx}{\textwidth}{l|*{6}{>{\centering\arraybackslash}X}}

\toprule
\toprule
&
\multicolumn{3}{c}{PRISM (R100)} &
\multicolumn{3}{c}{Grating (R1000)}\\

\cmidrule(lr){2-4}
\cmidrule(lr){5-7}

Compact \\ Population Lines &
MC pPXF Flux &
MC pPXF S/N &
Dust-corrected Flux  &
MC pPXF Flux  &
MC pPXF S/N &
Dust-corrected Flux \\


\midrule

Ly$\alpha$
& $1.36\pm0.25$ & 5.4 & $2.20\pm0.82$
& $2.29\pm0.23$ & 10.2 & $3.18\pm0.81$ \\

[N\,\textsc{iv}]$\lambda1483$
& -- & -- & -- 
& $<0.15$ & -- & $<0.19$ \\

N\,\textsc{iv}]$\lambda1486$*
& $0.25\pm0.08$ & 3.0 & $0.36\pm0.16$
& $<0.15$ & -- & $<0.19$ \\

C\,\textsc{iv}$\lambda1548$*
& $<0.21$ & -- & $<0.34$
& $<0.14$ & -- & $<0.18$  \\

C\,\textsc{iv}$\lambda1550$
& -- & -- & --
& $<0.14$ & -- & $<0.18$ \\

He\,\textsc{ii}$\lambda1640$
& -- & -- & --
& $0.12\pm0.03$ & 3.6 & $0.15\pm0.05$ \\

[O\,\textsc{iii}]$\lambda\lambda1661,1666$*
& $0.29\pm0.07$ & 4.0 & $0.41\pm0.15$
& $0.34\pm0.04$ & 7.6 & $0.42\pm0.10$ \\


N\,\textsc{iii}]$\lambda\lambda1747$-$1754$
& $<0.19$ & -- & $<0.26$
& $<0.45$ & -- & $<0.56$ \\

C\,\textsc{iii}]$\lambda1907$*
& $0.52\pm0.06$ & 9.4 & $0.70\pm0.18$
& $0.27\pm0.07$ & 4.1 & $0.33\pm0.10$ \\

C\,\textsc{iii}]$\lambda1909$
& -- & -- & --
& $0.21\pm0.06$ & 3.4 & $0.25\pm0.09$ \\



\midrule

[O\,\textsc{ii}]$\lambda3726$*
& $0.33\pm0.03$ & 10.9 & $0.35\pm0.06$
& $0.18\pm0.03$ & 5.4 & $0.18\pm0.04$ \\

[O\,\textsc{ii}]$\lambda3729$
& -- & -- & --
& $0.33\pm0.04$ & 8.9 & $0.35\pm0.06$ \\





[O\,\textsc{iii}]$\lambda4363$*
& $0.18\pm0.03$ & 6.6 & $0.18\pm0.04$
& $0.14\pm0.01$ & 11.3 & $0.15\pm0.02$ \\


H$\beta$
& $1.00\pm0.02$ & 43.8 & $1.00\pm0.10$
& $1.00\pm0.02$ & 48.6 & $1.00\pm0.07$ \\

[O\,\textsc{iii}]$\lambda\lambda5007,4959$
& $4.92\pm0.04$ & 128.0 & $4.87\pm0.54$
& $4.90\pm0.04$ & 118.4 & $4.88\pm0.41$ \\



H$\alpha$*
& $3.11\pm0.07$ & 47.0 & $2.86\pm0.28$
& $3.03\pm0.03$ & 87.8 & $2.86\pm0.20$ \\

[N\,\textsc{ii}]$\lambda\lambda6548,6584$*
& -- & -- & -- 
& $0.13\pm0.01$ & 9.4 & $0.12\pm0.01$ \\

[S\,\textsc{ii}]$\lambda6716$
& -- & -- & --
& $0.06\pm0.01$ & 6.0 & $0.06\pm0.01$ \\

[S\,\textsc{ii}]$\lambda6731$*
& $0.06\pm0.02$ & 3.2 & $0.06\pm0.02$
& $0.05\pm0.01$ & 4.5 & $0.04\pm0.01$ \\

\bottomrule
\bottomrule

\end{tabularx}

\begin{tablenotes}
\footnotesize
\item NOTE- * emission lines in the R100 that are blended at the native spectral resolution are fit simultaneously using single Gaussian model in \texttt{pPXF}. All fluxes and errors are reported from median and standard deviations of the 1000 Monte Carlo realizations. Summed errors upper limits correspond to $3\sigma$ non-detections. Dust-corrected fluxes are corrected using the Balmer decrement using the Calzetti Dust Attenuation Curve \citep{calzetti_dust_2000}.
\end{tablenotes}

\end{threeparttable}

\end{table*}


\begin{table*}
\centering
\begin{threeparttable}

\caption{Emission line fluxes normalized by H$\beta$ flux and S/N detections measured for the extended galaxies population in median PRISM R100 and medium resolution Grating R1000 spectra.}

\label{tab:linefluxes_extended}

\begin{tabularx}{\textwidth}{l|*{6}{>{\centering\arraybackslash}X}}

\toprule
\toprule
&
\multicolumn{3}{c}{PRISM (R100)} &
\multicolumn{3}{c}{Grating (R1000)}\\

\cmidrule(lr){2-4}
\cmidrule(lr){5-7}

Extended \\ Population Lines &
MC pPXF Flux &
MC pPXF S/N &
Dust-corrected Flux  &
MC pPXF Flux &
MC pPXF S/N &
Dust-corrected Flux \\

\midrule

Ly$\alpha$
& $<0.68$ & -- & $<1.63$
& $<1.51$ & -- & $<7.68$ \\

[N\,\textsc{iv}]$\lambda1483$
& -- & -- & -- 
& $<0.49$ & -- & $<1.77$ \\

N\,\textsc{iv}]$\lambda1486$*
& $<0.51$ & -- & $<1.02$
& $<0.48$ & -- & $<1.71$ \\

C\,\textsc{iv}$\lambda1548$*
& $<0.48$ & -- & $<0.91$
& $<0.43$ & -- & $<1.48$  \\

C\,\textsc{iv}$\lambda1550$
& -- & -- & --
& $0.66\pm0.18$ & 3.7 & $2.15\pm1.37$ \\

He\,\textsc{ii}$\lambda1640$
& -- & -- & --
& $<0.33$ & -- & $<1.05$ \\

[O\,\textsc{iii}]$\lambda\lambda1661,1666$*
& $0.63\pm0.16$ & 3.9 & $1.13\pm0.57$
& $<0.34$ & -- & $<1.06$ \\


N\,\textsc{iii}]$\lambda\lambda1747$-$1754$
& $<0.39$ & -- & $<0.69$
& $<1.54$ & -- & $<4.56$ \\

C\,\textsc{iii}]$\lambda1907$*
& $0.50\pm0.11$ & 4.4 & $0.83\pm0.36$
& $<0.29$ & -- & $<0.79$ \\

C\,\textsc{iii}]$\lambda1909$
& -- & -- & --
& $0.35\pm0.12$ & 3.0 & $0.90\pm0.56$ \\



\midrule

[O\,\textsc{ii}]$\lambda3726$*
& $0.53\pm0.07$ & 8.0 & $0.60\pm0.16$
& $0.73\pm0.14$ & 5.2 & $0.96\pm0.33$ \\

[O\,\textsc{ii}]$\lambda3729$
& -- & -- & --
& $0.60\pm0.14$ & 4.3 & $0.78\pm0.30$ \\





[O\,\textsc{iii}]$\lambda4363$*
& $0.25\pm0.07$ & 3.7 & $0.26\pm0.09$
& $0.13\pm0.03$ & 4.4 & $0.14\pm0.05$ \\


H$\beta$
& $1.00\pm0.04$ & 26.6 & $1.00\pm0.14$
& $1.00\pm0.05$ & 19.7 & $1.00\pm0.16$ \\

[O\,\textsc{iii}]$\lambda\lambda5007,4959$
& $5.12\pm0.06$ & 86.6 & $5.04\pm0.83$
& $5.20\pm0.12$ & 44.86 & $5.03\pm1.02$ \\



H$\alpha$
& $3.32\pm0.08$ & 40.2 & $2.86\pm0.40$
& $3.80\pm0.08$ & 44.9 & $2.86\pm0.46$ \\

[N\,\textsc{ii}]$\lambda\lambda6548,6584$*
& -- & -- & --
& $0.13\pm0.03$ & 4.2 & $0.10\pm0.03$ \\

[S\,\textsc{ii}]$\lambda6716$
& -- & -- & --
& $0.13\pm0.03$ & 5.1 & $0.10\pm0.02$ \\

[S\,\textsc{ii}]$\lambda6731$*
& $0.21\pm0.04$ & 5.2 & $0.18\pm0.04$
& $0.11\pm0.02$ & 5.1 & $0.08\pm0.02$ \\

\bottomrule
\bottomrule

\end{tabularx}

\begin{tablenotes}
\footnotesize
\item NOTE- * emission lines in the R100 that are blended at the native spectral resolution are fit simultaneously using single Gaussian model in \texttt{pPXF}. All fluxes and errors are reported from median and standard deviations of the 1000 Monte Carlo realizations. Summed errors upper limits correspond to $3\sigma$ non-detections. Dust-corrected fluxes are corrected using the Balmer decrement using the Calzetti Dust Attenuation Curve \citep{calzetti_dust_2000}.
\end{tablenotes}

\end{threeparttable}
\end{table*}

\subsection{Emission Line Fitting}
\label{sec:ppxf}

We perform emission line fitting for the presented stacked spectra  using the Penalized PiXel-Fitting code (\texttt{pPXF}; \citealp{cappellari_parametric_2004,cappellari_improving_2017}). The fitting is independently performed over the rest-frame UV ($1150$-$3000$~\AA) \ and rest-frame optical ($3000$-$7000$~\AA) wavelength ranges. Although the R1000 observations provide substantially higher spectral resolution than the R100 prism spectra, their exposure times are typically shorter by a factor of $\sim1$-$4$ \citep{deugenio_jades_2024}. For the R1000 stacks we further split the rest-frame optical fit into three local windows ($3000$--$4000$~\AA, $4000$--$4600$~\AA\ and $4600$--$7000$~\AA) , so that
any residual systematics in one region from spectral overlapping do not bias the fit elsewhere. Consequently, we use both datasets depending on the wavelength coverage and the S/N of the emission lines of interest. The higher S/N of the R100 stacks enables reliable measurements of the weaker UV emission lines, including \textsc{N~iv}]~$\lambda\lambda1486,1483$, He~\textsc{ii}~$\lambda1640$, \textsc{O~iii}]~$\lambda\lambda1661,1666$, \textsc{N~iii}]~$\lambda\lambda1746$-1754, and \textsc{C~iii}]~$\lambda\lambda1907,1909$. In contrast, the higher spectral resolution of the R1000 observations allows several important line blends to be resolved, including [O~\textsc{ii}]~$\lambda\lambda3726,3729$, [O~\textsc{iii}]~$\lambda4363$, [O~\textsc{iii}]~$\lambda\lambda4959,5007$, H$\beta$, H$\alpha$, [N~\textsc{ii}]~$\lambda\lambda6548,6584$, and [S~\textsc{ii}]~$\lambda\lambda6716,6731$.

\begin{figure}
	
	\includegraphics[width=1.0\columnwidth]{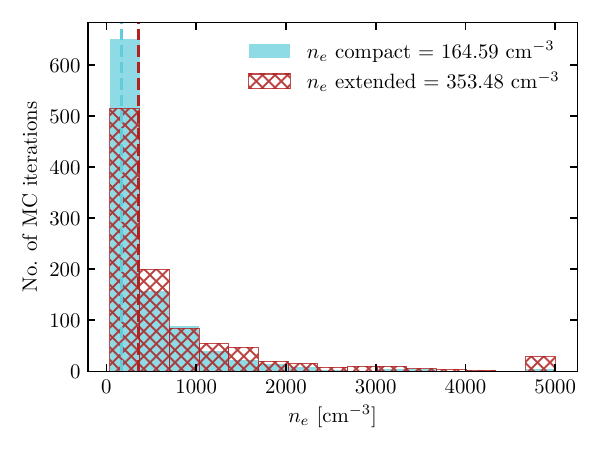}
    \caption{The $n_{\rm e}$ estimates calculated using the [S~\textsc{ii}]$\lambda6716/\lambda6731$ line ratio from the rest-frame optical stacks using \texttt{PyNeb}. This shows a distribution of the electron densities, $n_{\rm e}$, based on the distribution of fluxes calculated with the 1000 MC iterations performed on the grating stack in pPXF. The blue represents the $n_{\rm e}$ estimates calculated from the compact grating stack and the red for the $n_{\rm e}$ extended stack. The median values of each of the distributions are shown via the dashed line. We obtain a median value of $n_{\rm e, compact} \sim 233.14 \ \rm cm^{-3}$ for the compact stack and $n_{\rm e, extended} \sim 337.31 \ \rm cm^{-3}$ for the extended stack.}
    
    \label{fig:ne_dist}
\end{figure}

We use pPXF to extract the gas kinematics of our key emission lines. \texttt{pPXF} uses a linearly constrained non-linear optimization program which uses Sequential Quadratic Programming (SQP) and Levenberg-Marquardt (LM) methods. The LM solves the non-linear parameters which are generally the stellar kinematic components like velocity and velocity dispersion using weights which are used for normalization. The SQP solves the fit by treating it as a quadratic problem with given constraints.

For each wavelength region, we adopt the wavelength-dependent \textit{nominal} NIRSpec prism and grating instrumental resolution for the R100 and R1000 spectra respectively \citep{jakobsen_near-infrared_2022}. The stacked spectra are modelled using two components. The first component represents the underlying continuum, which is described using a combination of high-order additive and multiplicative polynomials together with power-law continua where appropriate. The second component consists of Gaussian emission line templates corresponding to the transitions being fitted. The continuum and emission line templates are fitted simultaneously, with each component having independent kinematics described by two moments: velocity  and velocity dispersion. Throughout this fitting, we adopt additive and multiplicative polynomial degrees of two.

The Gaussian amplitudes returned by \texttt{pPXF} are converted into integrated emission line fluxes using the corresponding wavelength-dependent Gaussian widths. To estimate robust fluxes and their uncertainties, we perform 1000 Monte Carlo realizations of each stacked spectrum. For each trial, the stacked spectrum is perturbed according to is $1\sigma$ uncertainty assuming Gaussian statistics, and the complete \texttt{pPXF} fitting procedure is repeated. The resulting distributions of integrated line fluxes are used throughout this work to derive emission line measurements and their associated uncertainties. We set a $S/N > 3$ threshold for emission line detections and report the $3\sigma$ upper limits using summed errors for the undetected emission line transitions. 

In the UV wavelength region, we detect several key diagnostic emission lines in the stacked spectra. The compact median R100 stack exhibits a $\sim3.0\ \sigma$ detection of N~\textsc{iv}], whereas this line is not significantly detected in the extended population. We also find significant detections of C~\textsc{iii}]$\lambda\lambda1907,1909$, O~\textsc{iii}]$\lambda\lambda1661,1666$ in both morphology populations in median R100 stacks with $S/N>3$. We also detect distinct differences in the C~\textsc{iv} profile between the two samples: the compact stack displays a non-gaussian C~\textsc{iv} doublet profile, visually exhibiting emission features, while the extended stacks exhibit predominately absorption. To prevent the complex C~\textsc{iv} profile from biasing the neighbouring N~\textsc{iv}] kinematics measurements, we fit the C~\textsc{iv} doublet independently with its own kinematic component.



The optical stacked R100 and R1000 spectra exhibit strong detections of the Balmer emission lines H$\alpha$ and H$\beta$, together with the principal oxygen diagnostics, including [O~\textsc{ii}]~$\lambda\lambda3726,3729$, [O~\textsc{iii}]~$\lambda4363$, and [O~\textsc{iii}]~$\lambda\lambda5007,4959$. We also detect the [S~\textsc{ii}]~$\lambda\lambda6716,6731$ doublet with high significance. Even though these lines are detected in the R100 stacks, they are further well-resolved in the R1000 stacks. Although the [O~\textsc{iii}] $\lambda\lambda5007,4959$ doublet is fully resolved in the R1000 spectra, \texttt{pPXF} constrains the relative line strengths of
the two transitions according to their theoretical transition probabilities,
with a ratio of $2.98$ under optically thin nebular conditions
\citep{dufresne_chianti_2024}. The resulting emission-line measurements form the basis of the physical diagnostics presented in the following sections.

\section{Nebular Properties} \label{sec:neb_prop}

In this section, we discuss the nebular properties derived from the stacked spectra fits. 

\subsection{Direct Method $T_{\rm e}$-based Oxygen Abundances}

We measure the gas-phase oxygen abundances of the stacked spectra using the electron temperature "\textit{direct} method", which constrains the electron temperature, $T_{\rm e}$, through the temperature of sensitive auroral emission lines. In particular, we derive $T_{\rm e}$ from the flux ratio [O~\textsc{iii}]~$\lambda\lambda5007,4959$/[O~\textsc{iii}]~$\lambda4363$. Since the stacked R1000 spectra resolve [O~\textsc{iii}]~$\lambda4363$ from the neighbouring H$\gamma$ transition, we are able to obtain reliable flux measurements from the \texttt{pPXF} fits for both the compact and extended galaxy populations.

We calculate the electron temperature $T_{\rm e}$, electron density $n_{\rm e}$, gas-phase oxygen abundances (or metallicities, $Z$), and ionic abundances using the \texttt{PyNeb} package, which computes emission line emissivities by solving the equilibrium equations for the relevant atomic species \citep{luridiana_pyneb_2015}. We adopt the atomic data compilation from \citet{isobe_empress_2023}, while using the collisional strengths for $\rm O^{2+}$ from \citet{lennon_atomic_1994}.

Before deriving the nebular properties, we correct the stacked spectra for dust attenuation using the \citet{calzetti_dust_2000} attenuation law, which was derived for local star-forming galaxies and is commonly adopted for high-redshift star-forming galaxies. 
We inferred the dust attenuation from the $H\alpha/H\beta$ Balmer decrement, assuming an intrinsic case B value of 2.86. The intrinsic ratio varies by only a few per cent over the range of temperatures and densities expected in H~\textsc{ii} regions. This is small compared with the uncertainties on our stacked Balmer decrements, so we adopt a single value to treat both stacks consistently.




\begin{figure}
	
	\includegraphics[width=1.0\columnwidth]{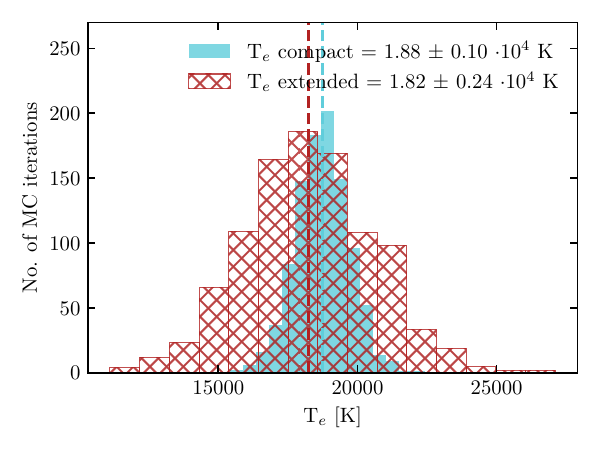}
    \caption{The direct-method $T_{\rm e}$ estimates calculated using the [OIII]$\lambda5007, 4959$/[OIII]$\lambda4363$ line ratio from the rest-frame optical stacks using \texttt{PyNeb}. This shows a distribution of the temperatures, $T_{\rm e}$, based on the distribution of fluxes calculated with the 1000 MC iterations performed on the grating stack in \texttt{pPXF}. The blue represents the $T_{\rm e}$ estimates calculated from the compact grating stack and the red for the $T_{\rm e}$ extended stack. The median values of each of the distributions are shown via in the dashed line. We obtain a median value of $T_{\rm e, \ compact} = 1.88 \pm 0.10 \times 10^4K$ for the compact stack and $T_{\rm e, \ extended} = 1.82 \pm 0.24 \times 10^4 K$ for the extended stack.}
    \label{fig:Te_dist}
\end{figure}

The electron density, $n_{\rm e}$, is determined from the [S~\textsc{ii}]$\lambda6716/\lambda6731$ doublet ratio using \texttt{PyNeb}, assuming an electron temperature of $T_{\rm e}=10^4$~K. The ratio depends only weakly on temperature, so this assumption has a negligible effect on the derived densities. Both the [S~\textsc{ii}]$\lambda6716$ and [S~\textsc{ii}]$\lambda6731$ lines are detected in the R1000 stacked spectra fits which yield electron densities of $n_{\rm e, \ compact} \sim 164.59 \ \rm{cm}^{-3}$ and $n_{\rm e, \ extended} \sim 353.48 \ \rm{cm}^{-3}$. 

We account for the low and high density limits when determining $n_{\rm e}$. In some MC iterations for the flux ratios, the measured [S~\textsc{ii}] ratio lies outside  the range over which \texttt{PyNeb} provides a unique density solution, either exceeding the low-density limit ($\lambda6716/\lambda6731 \gtrsim 1.45$) or falling below the high-density limit ($\lambda6716/\lambda6731 \lesssim 0.59$). We therefore adopt $n_{\rm e}=40~\rm{cm}^{-3}$ and $n_{\rm e}=5000~\rm{cm}^{-3}$ as the low- and high-density limits, respectively, following \citet{kewley_theoretical_2019}. Using these bounds, we determine the range of possible $n_{\rm e}$ values across the MC-perturbed flux ratios derived from the \texttt{pPXF} fits. 

The resulting distributions of $n_{\rm e}$ are shown in Figure~\ref{fig:ne_dist}, with the corresponding values reported in Table~\ref{tab:nebular_prop}. Most MC realizations lie within the low-density regime. Given the large measurement uncertainties in both the compact and extended samples, we propagate the full distribution of MC-derived $n_{\rm e}$ values through the subsequent calculations, thereby accounting for the uncertainty in the electron density.

These densities are low compared with those inferred for extreme compact, N-enhanced systems such as GN-z11 at $z=10.6$, where the UV nitrogen lines imply $n_{\rm e}\gtrsim10^{4-5}$~cm$^{-3}$ \citep{senchyna_gn-z11_2023, maiolino_small_2024}.
However, the two diagnostics trace different gas. The UV N\,\textsc{iv}] and C\,\textsc{iii}] lines probe the dense, high-ionization zone, while [S\,\textsc{ii}] traces low-ionization gas and saturates above $\sim10^{3.5}$~cm$^{-3}$. Moreover, the N/O--$n_{\rm e}$([S\,\textsc{ii}]) correlation reported at $z\sim0$ \citet{arellano-cordova_classy_2025} remains debated, whereas N/O correlates consistently with $\Sigma_{\rm SFR}$ \citep{cameron_jades_2026}. $\Sigma_{\rm SFR}$ traces galaxy-scale star formation whereas
$n_{\rm e}$([S\,\textsc{ii}]) reflects local H\,\textsc{ii}-region conditions shaped by feedback. Further, the MEGATRON simulations from \citep{choustikov_megatron_2026} predict lower $n_{\rm e}$([S\,\textsc{ii}]) for bursty, supernova-driven star formation than for efficient star formation. A low $n_{\rm e}$([S\,\textsc{ii}]) therefore does not rule out compact, N-enriched star formation.

The electron temperature is derived from the dust-corrected [O~\textsc{iii}]~$\lambda\lambda5007,4959$/[O~\textsc{iii}]~$\lambda4363$ ratio using \texttt{PyNeb}. The auroral line, [O~\textsc{iii}]~$\lambda4363$, is detected in both the morphology populations which yields $T_{\rm e}$ values in the range of $12000-20000 \ \rm{K}$ as shown in Figure~\ref{fig:Te_dist}. These temperatures are then used to compute the ionic abundances through the corresponding nebular emissivities using the relevant strong emission lines from the rest-frame UV. 

We also calculate the gas-phase oxygen abundances (a.k.a. metallicities) using these $T_{\rm e}$(\rm[O~\textsc{iii}]) and $n_{\rm e}$(\rm[S~\textsc{ii}]) estimates. The ion abundance ratio of $\rm O^{2+}/H^+$ is derived from [O~\textsc{iii}]~$\lambda5007,4959$/H$\beta$ using the directly measured temperature, $T_e(\rm O^{2+})$. The abundance of $\rm O^+/H^+$ is derived from [O~\textsc{ii}]~$\lambda\lambda3726,3729$/H$\beta$ ratio adopting \begin{equation}\label{eq:TeO2} 
    T_{\rm e}({\rm[OII]}) = (0.54\pm0.03) \ T_{\rm e} + (4790\pm300)\ \rm K
\end{equation}
from \citet{cataldi_marta_2025} following the adapted two-zone temperature approximation. The ionic abundances: $\rm O^{+}/H^{+}$ and $\rm O^{2+}/H^{+}$ are calculated as

\begin{equation}\label{eq:O+/H+} 
    {\rm O^+ \over {H^+}} = {I_{\lambda\lambda3726,3729} \over I_{\rm H\beta}} \times {\epsilon_{\rm H\beta} \over \epsilon_{\lambda\lambda3726,3729}}
\end{equation}

\begin{equation}\label{eq:O2+/H+} 
    {\rm O^{2+} \over {H^+}} = {I_{\lambda\lambda5007,4959} \over I_{\rm H\beta}} \times {\epsilon_{\rm H\beta} \over \epsilon_{\lambda\lambda5007,4959}}
\end{equation}
where $\epsilon$ denotes the emissivity of the corresponding transition calculated by \texttt{PyNeb}.

The total gas-phase oxygen abundance is obtained by summing the singly and doubly ionized oxygen abundances, 

\begin{equation}\label{eq:Z} 
    Z = 12 + \log_{10}\left(\frac{\mathrm{O}^{+} + \mathrm{O}^{2+}}{\mathrm{H}^{+}}\right)
\end{equation}
These give us the metallicities of $12+\log(\rm O/H)_{\rm compact} = 7.44\pm0.05$ for the compact and $12+\log(\rm O/H)_{\rm extended} = 7.62\pm0.13$ for the extended galaxy populations. The derived electron densities, electron temperatures, and metallicities are summarized in Table~\ref{tab:nebular_prop}.

        




\subsection{Nitrogen Abundance Measurement}

We use the electron temperatures, $T_{\rm e}$([O~\textsc{iii}]), electron densities, $n_{\rm e}$([S~\textsc{ii}]), and metallicities, $Z$, derived from the previous section to determine the nitrogen abundances from the median R100 and R1000 stacked spectra. Our analysis considers both the rest-frame UV and optical nitrogen emission lines, which probe the different ionisation states of nitrogen, including $\rm N^{+}$, $\rm N^{2+}$ and $\rm N^{3+}$. This allows us to calculate independent estimates of log(N/O) from the UV and optical spectra.

\subsubsection{Optical log(N/O) Measurement}

The optical [N~\textsc{ii}]~$\lambda6548,6584$ doublet line is commonly used to derive nitrogen abundances. Since these lines are blended with the broad $\rm H\alpha$ emission in the R100 stacked spectra, we instead use the R1000 stacked spectra to derive the optical nitrogen abundances. The [N~\textsc{ii}]~$\lambda6548,6584$ line is detected with $S/N\approx10$ in the compact R1000 median stack and $S/N\approx5$ in the extended R1000 median stack. We also detect [O~\textsc{ii}]~$\lambda\lambda3726,3729$ lines in both the morphology stacks in the R1000 observations, allowing us to estimate the nitrogen-to-oxygen abundance ratio from the ionic abundance ratio, $\rm log(N/O)_{optical} = N^+/O^+$.

The ionic abundance ratio $\rm N^{+}/O^{+}$ is derived from the [N~\textsc{ii}]$\lambda\lambda6548,6584$/[O~\textsc{ii}]$\lambda\lambda3726,3729$ line ratio from rest-frame optical. The ionic abundance ratios are calculated according to

\begin{equation}\label{eq:N+/O+} 
    {\rm N^{+} \over {O^{+}}} = {I_{\lambda\lambda6548,6584} \over I_{\lambda\lambda3726,3729}} \times {\epsilon_{\lambda\lambda3726,3729} \over \epsilon_{\lambda\lambda6548,6584}}
\end{equation}

Since the abundance calculation is based on the [O~\textsc{ii}] doublet, we adopt the $T_{\rm e} (\rm[OII])$ estimates. We estimate this temperature using the two-zone temperature approximation from \citet{cataldi_marta_2025} shown in Equation \ref{eq:TeO2}.

To account for the unobserved ionisation states of nitrogen, we apply the ionisation correction factors (ICFs) derived from the \texttt{CLOUDY} \citep[c17.03][]{ferland_2017_2023} photoionisation models presented by \citet{martinez_under_2026}. We adopt the calibration corresponding to the low-density regime $n_{\rm e} = 100 \ \rm cm^{-3}$, which is appropriate for electron densities derived for our stacked spectra (Table~\ref{tab:nebular_prop}). The ICF first requires an estimate of the ionisation parameter ${\rm log}(U)$, which is determined from the strong-line ratio $\rm O_{32}=[\frac{[\mathrm{O,III}]~\lambda\lambda4959,5007}{[\mathrm{O,II}]~\lambda\lambda3726,3729}
]$, together with the metallicity calculated from the R1000 stacks.

The ionisation parameter is estimated from the polynomial calibration
\begin{equation}
\begin{aligned}
f(x,y) = \log(U) =\;& A + Bx + Cy + Dxy + Ex^2 + Fy^2 \\
&+ Gxy^2 + Hyx^2 + Ix^3 + Jy^3
\end{aligned}
\end{equation}
where the $x =\rm log(O_{32})$ and $y=Z/Z_{\odot}$. The polynomial coefficients are given in \citet{martinez_under_2026}. 

Using this calibration, we obtain ${\rm log}(U)_{\rm compact}= -2.21\pm0.03$ for the compact stack and ${\rm log}(U)_{\rm extended}= -2.70\pm0.04$ for the extended stack. These ionisation parameter values are then used to evaluate the nitrogen ICFs from \texttt{CLOUDY} photoionisation models performed in \citet{martinez_under_2026}, using the following equation:

\begin{equation}
\begin{aligned}
f(x,y) = \rm ICF \left({\rm N^+\over O^+}\right)  =\;& A + Bx + Cy + Dxy + Ex^2 + Fy^2 \\
&+ Gxy^2 + Hyx^2 + Ix^3 + Jy^3
\end{aligned}
\end{equation}
where the $x ={\rm log}(U)$ and $y=Z/Z_{\odot}$.  The polynomial
coefficients are given in \citet{martinez_under_2026}. We obtain ICFs of $\sim1.00$ and $\sim1.04$ for the compact and extended stacks, respectively, which are subsequently applied to the ionic abundance ratios.

The total optical nitrogen-to-oxygen abundance ratio is then computed as

\begin{equation}\label{eq:N/O_optical} 
    \rm log\left({\rm N \over {O}}\right)_{optical}= \rm log\left(\rm {N^{+} \over O^{+}} \times \rm ICF({N^{+}/ O^{+})}\right)
\end{equation}

Using this approach, we get $\rm log(N/O)_{optical} = -0.72 \pm 0.05$ for the compact galaxy population and $\rm log(N/O)_{optical} = -1.27 \pm 0.11$ for the extended population. The compact galaxies therefore exhibit significantly enhanced nitrogen abundances relative to the extended population. The inferred abundance ratio for the compact stack also exceeds the solar nitrogen-to-oxygen abundance ratio as shown in Figure~\ref{fig:log(N/O)_median}.

\begin{table}
	\centering
	\caption{Median nebular properties of both morphology populations.}
	\label{tab:nebular_prop}
	\begin{tabular}{l|cc} 
		\hline
		Property & Compact & Extended \\
        \hline \hline
		$\rm H\alpha /H\beta$ & $3.02\pm0.07$ & $3.79\pm0.21$ \\
		${\rm log} (n_{\rm e}) \rm \ [cm^{-3}]$ & $2.22\pm0.97$ & $2.55\pm0.76$ \\
		$T_e \rm \ [10^4 \ K]$ & $1.88\pm0.10$ & $1.82\pm0.24$ \\
        $12+\rm log(O/H)$ & $7.44\pm0.05$ & $7.62\pm0.13$ \\
        ${\rm log}(U)$ & $-2.21\pm0.03$ & $-2.70\pm0.04$ \\
        $Z/Z_\odot$ & $0.06\pm0.01$ & $0.09\pm0.03$ \\
        ICF($\rm N^{+}/O^{+}$) & $1.00\pm0.004$ & $1.04\pm0.007$ \\
        ICF($\rm N^{2+}+N^{3+}/O^{2+}$) & $0.99\pm0.002$ & $0.93\pm0.005$ \\
        ICF($\rm C^{2+}/O^{2+}$) & $1.04\pm0.01$ & $0.86\pm0.02$ \\
        $\rm log(N/O)_{optical}$ & $-0.72\pm0.05$ & $-1.27\pm0.11$ \\
        $\rm log(N/O)_{UV, \ low \ density}$ & $>-0.69$ & $<-0.43$ \\
        $\rm log(N/O)_{UV, \ high \ density}$ & $>-0.69$ & $<-0.51$ \\
        log(C/O) & $-0.58\pm0.11$ & $-1.02\pm0.14$ \\

		\hline \hline
	\end{tabular}
\end{table}

\subsubsection{UV log(N/O) Measurement}

We derive nitrogen abundances from the rest-frame UV transitions [N~\textsc{iv}]~$\lambda\lambda1483,1486$ and N~\textsc{iii}]~$\lambda1746$-$1754$, which trace more highly ionized gas than the optical lines. We detect [N~\textsc{iv}] with a $S/N=3.0$ in only the compact R100 stack, while we do not detect the N~\textsc{iii}] line in either and instead use $3\sigma$ upper limits for N~\textsc{iii}] fluxes. Owing to the spectral resolution of the PRISM observations, any broad and narrow components cannot be separated, and we therefore model the UV emission lines using a single broad component \citep{labbe_unambiguous_2024}. 

Since we detect the [N~\textsc{iv}] line the compact sources, we calculate the ionic abundance ratio $\rm N^{3+}/O^{2+}$ from [N~\textsc{iv}]~$\lambda\lambda1483,1486$/O~\textsc{iii}]$\lambda\lambda1661,1666$ line ratio. Since the O~\textsc{iii}]$\lambda\lambda1661,1666$ line is blended with the He~\textsc{ii}]$\lambda1640$ line in the R100 stacked spectra, the derived nitrogen abundance ratios are shown with the blended O~\textsc{iii}] + He~\textsc{ii} fluxes. The \citet{martinez_under_2026} ICF calibration yields non-physical large correction factors for this ionisation state which are $\geq 1$. Consequently, we do not apply an ICF and instead report a lower limit on $\log({\rm N/O})$ for the compact stack. We therefore estimate $\rm log(N/O)_{UV, \ compact} = N^{3+}/O^{2+}$ which is calculated according to

\begin{equation}\label{eq:N3+/O2+_compact} 
    {\rm N^{3+} \over {O^{2+}}} = {I_{\lambda\lambda1486,1483} \over I_{\lambda\lambda1661,1666}} \times {\epsilon_{\lambda\lambda1661,1666} \over \epsilon_{\lambda\lambda1486,1483}}
\end{equation}
where the $I_\lambda$ denotes the integrated emission-line fluxes measured from the Monte Carlo \texttt{pPXF} and the $\epsilon_\lambda$ represent the corresponding emissivities calculated from \texttt{PyNeb}. The total nitrogen-to-oxygen abundance ratio is then computed as

\begin{equation}\label{eq:N/O_UV_compact} 
    \rm log\left({\rm N \over {O}}\right)_{UV, \ compact}>\rm log\left(\rm {N^{3+} \over O^{2+}}\right)
\end{equation}
Using this approach, we get $\rm log(N/O)_{UV, \ optical} > -0.69$ for the compact galaxy population.

Since neither [N~\textsc{iv}]~$\lambda\lambda1483,1486$ nor the N~\textsc{iii}]~$\lambda1746$-$1754$ is detected in the extended stack, we instead derive an upper limit on $\log[\rm (N^{2+}+N^{3+})/O^{2+}]$.
The ionic abundance ratio $\rm N^{3+}/O^{2+}$ is derived from the [N~\textsc{iv}]~$\lambda\lambda1483,1486$/O~\textsc{iii}]$\lambda\lambda1661,1666$ line ratio and $\rm N^{2+}/O^{2+}$ using N~\textsc{iii}]$\lambda1746$-$1754$/O~\textsc{iii}]$\lambda\lambda1661,1666$ ratio from rest-frame UV. We therefore estimate  $\rm log(N/O)_{UV, \ extended} = (N^{3+}+N^{2+})/O^{2+}$. The ionic abundance ratios are calculated according to

\begin{equation}\label{eq:N3+/O2+_extended} 
    {\rm N^{3+} \over {O^{2+}}} = {I_{\lambda\lambda1486,1483,3\sigma} \over I_{\lambda\lambda1661,1666}} \times {\epsilon_{\lambda\lambda1661,1666} \over \epsilon_{\lambda\lambda1486,1483}}
\end{equation}

\begin{equation}\label{eq:N2+/O2+} 
    {\rm N^{2+} \over {O^{2+}}} = {I_{\lambda\lambda1747\text{-}1754,3\sigma} \over I_{\lambda\lambda1661,1666}} \times {\epsilon_{\lambda\lambda1661,1666} \over \epsilon_{\lambda\lambda1747\text{-}1754}}
\end{equation}
where the $I_{\lambda, 3\sigma}$ denotes the integrated emission line fluxes measured from the Monte Carlo \texttt{pPXF} and the $\epsilon_\lambda$ represent the corresponding emissivities calculated from \texttt{PyNeb}.

We have ${\rm log}(U)_{\rm extended}= -2.70 \pm 0.04$ for the extended stack calculated from the previous section. These ionisation parameter values are then used to calculate the ICFs for the nitrogen abundances from \texttt{CLOUDY} photoionisation models performed in \citet{martinez_under_2026}, using the following equation:

\begin{equation}
\begin{aligned}
f(x,y) = \rm ICF \left({\rm N^{2+} + N^{3+} \over O^{2+}}\right) =\;& A + Bx + Cy + Dxy + Ex^2 + Fy^2 \\
&+ Gxy^2 + Hyx^2 + Ix^3 + Jy^3
\end{aligned}
\end{equation}
where the $x ={\rm log}(U)$ and $y=Z/Z_{\odot}$. The polynomial
coefficients are given in \citet{martinez_under_2026}. We obtain ICF values of $\sim0.93$ for the extended stack. We then apply this correction to our extended population abundance calculations. The total nitrogen-to-oxygen abundance ratio for the extended population is then computed as

\begin{equation}\label{eq:N/O_UV_extended} 
    \rm log\left({\rm N \over {O}}\right)_{UV, \ extended} <\rm log\left(\rm {N^{2+}_{3\sigma}+ N^{3+}_{3\sigma} \over O^{2+}}\times \rm ICF({(N^{2+} + N^{3+})/ O^{2+}})\right) 
\end{equation}

We obtain $\rm log(N/O)_{UV, \ extended} < -0.43$ for the extended population. These values are shown in the Figure~\ref{fig:log(N/O)_median}. 

To further compare the UV measurements in high density regime, we assume a density of $n_{\rm e} =10^5 \rm \ cm^{-3}$ to take into account the possible high density values expected from high ionization UV lines, since these are not resolved in the low resolution R100 prism spectra and not detected in the R1000 grating spectra. With this assumption we get, $\rm log(N/O)_{UV, \ compact} <-0.69$ for the compact population and $\rm log(N/O)_{UV, \ extended} <-0.43$ for the extended population. These values are shown in the Figure~\ref{fig:log(N/O)_median}. 

To further compare the UV-based measurements in the high-density regime, we adopt an electron density of $n_{\rm e} = 10^5\ {\rm cm^{-3}}$ \citep{senchyna_gn-z11_2023, ji_connecting_2025}. High-ionization UV lines are expected to trace such high densities, but the density-sensitive doublets are unresolved in the low-resolution R100 prism spectra and undetected in the R1000 grating spectra, so we cannot measure $n_{\rm e}$ directly. Under this assumption, we obtain upper limits of $\log({\rm N/O})_{\rm UV,\,compact} <-0.69$ for the compact population and $\log({\rm N/O})_{\rm UV,\,extended} <-0.51$ for the extended population, as shown in Table~\ref{tab:nebular_prop}.

\subsection{Carbon Abundance Measurement}

We also detect C~\textsc{iii}] emission in the rest-frame UV stacked spectra in both galaxy populations. In contrast, the C~\textsc{iv}~$\lambda\lambda1548,1551$ doublet exhibits a more complex profile. The compact R100 stack shows an emission component, while the extended stack is dominated by absorption. Owing to the resonant nature of the C~\textsc{iv} transition, we fit the kinematics of this line independently from the other UV lines. However, C~\textsc{iv} is significantly detected only in the compact stack and is therefore not used in our abundance analysis. Hence, we derive the carbon abundance using the non-resonant C~\textsc{iii}]~$\lambda\lambda1907,1909$ doublet. Throughout this work, we assume that the elemental abundance ratio is well approximated by $\rm log(C/O) = C^{2+}/O^{2+}$ after applying the appropriate ionisation correction factor (ICF).

We use the electron temperatures, $T_{\rm e}$(\rm[O~\textsc{iii}]), electron densities, $n_{\rm e}$(\rm[S~\textsc{ii}]), and metallicities, $Z$, estimates derived in the previous section to calculate the carbon abundances. To account for unobserved ionisation states, we adopt the ICF calibrations based on the \texttt{CLOUDY} photoionisation models by \citet{martinez_under_2026, berg_chemical_2019}. Since the ionisation parameters have already been determined as shown in Table~\ref{tab:nebular_prop}, the carbon ICFs can be calculated directly using the polynomial calibration from \citet{berg_chemical_2019}.
\begin{equation}
    f(x) = {\rm ICF \left({ C^{2+}\over O^{2+}}\right)}=Ax^3+Bx^2+Cx+D.
\end{equation}
where $x = \rm log(U)$ and the coefficients are listed in \citet{berg_chemical_2019}. 

The metallicities derived from the R1000 stacks correspond to approximately $Z= 0.05 Z_{\odot}$ for the compact population and $Z= 0.09 Z_{\odot}$ for the extended population. Since the \texttt{CLOUDY} model fits are tabulated for $0.05$, $0.1$, $0.2$, and $0.4,Z_\odot$ in \citet{berg_chemical_2019}, we use the 0.05$Z_\odot$ calibration for the compact stack and 0.1$Z_\odot$ calibration for the extended stack. This yields ICF values for $\rm C^{2+}/O^{2+}$ of approximately $\sim1.05$ and $\sim~0.86$ for the compact and extended populations respectively. 

The ionic abundance ratio $\rm C^{2+}/O^{2+}$ is calculated from the C~\textsc{iii}]~$\lambda\lambda1907,1909$/O~\textsc{iii}]~$\lambda\lambda1661,1666$ line ratio according to

\begin{equation}\label{eq:C2+/O2+} 
    {\rm C^{2+} \over {O^{2+}}} = {I_{\lambda\lambda1907.1910} \over I_{\lambda\lambda1661,1666}} \times {\epsilon_{\lambda\lambda1661,1666} \over \epsilon_{\lambda\lambda1907,1910}}
\end{equation}
where $I_\lambda$ denotes the integrated emission-line fluxes measured from the Monte Carlo \texttt{pPXF} fits and $\epsilon_\lambda$ represents the corresponding emissivities calculated using \texttt{PyNeb}. As discussed in the previous section, the O~\textsc{iii}]~$\lambda\lambda1661,1666$ doublet is partially blended with He~\textsc{ii}~$\lambda1640$, and therefore the derived carbon abundance ratios are
shown with the blended O~\textsc{iii}] + He~\textsc{ii} fluxes.

The elemental carbon-to-oxygen abundance ratio is then calculated as
\begin{equation}\label{eq:C/O} 
    \rm log \left({\rm C \over {O}}\right)= \rm log \left(\rm {C^{2+} \over O^{2+}} \times \rm ICF({C^{2+}/ O^{2+}})\right)
\end{equation}
Using this approach, we derive $\rm log(C/O) = -0.58 \pm 0.11$ for the compact galaxy population and $\rm log(C/O) = -1.02 \pm 0.14$ for the extended galaxy population. The compact galaxies exhibit enhanced carbon abundances relative to the extended galaxies as shown in Figure~\ref{fig:log(C/O)_median}.

\section{Results and Discussion}\label{sec:results_disc}

\subsection{Elevated N/O}

We detect the [N~\textsc{ii}]$\lambda\lambda6748,6784$ in the median R1000 stacked spectra of both galaxy populations. We also detect the N~\textsc{iv}]$\lambda1483,1486$ emission line in the median R100 stacked spectra of the compact galaxy population, whereas this feature is not significantly detected in the extended stacks. As mentioned previously, the R1000 spectra have substantially lower exposure times than the R100 observations, likely resulting in insufficient S/N for the detection of these intrinsically weak UV nitrogen transitions, even after stacking. 

To quantify the chemical differences between the two morphological galaxy populations, we calculate nitrogen ionic abundances from the optical [N~\textsc{ii}] lines and UV N~\textsc{iv}] lines. We compare these measurements with a compilation of the known nitrogen-enhanced galaxies (NOEGs) from the literature \citep{castellano_jwst_2024, cameron_nitrogen_2023, cameron_nebular_2024, schaerer_discovery_2024, isobe_jwst_2023, larson_ceers_2023, topping_deep_2024, topping_aurora_2025, ji_ga-nifs_2024, ji_connecting_2025, ubler_ga-nifs_2023,pascale_nitrogen-enriched_2023, james_vlt_2009, arellano-cordova_classy_2025, napolitano_seven_2025, navarro-carrera_constraints_2024, stiavelli_what_2025, zhang_potential_2025}, Milky Way globular clusters and field stars from APOGEE \citep{abdurrouf_seventeenth_2022} and nearby metal-poor galaxies \citep{izotov_chemical_2006, pilyugin_abundance_2012, annibali_chemical_2019, grossi_southern_2025}. 

\begin{figure}
	\includegraphics[width=1\columnwidth]{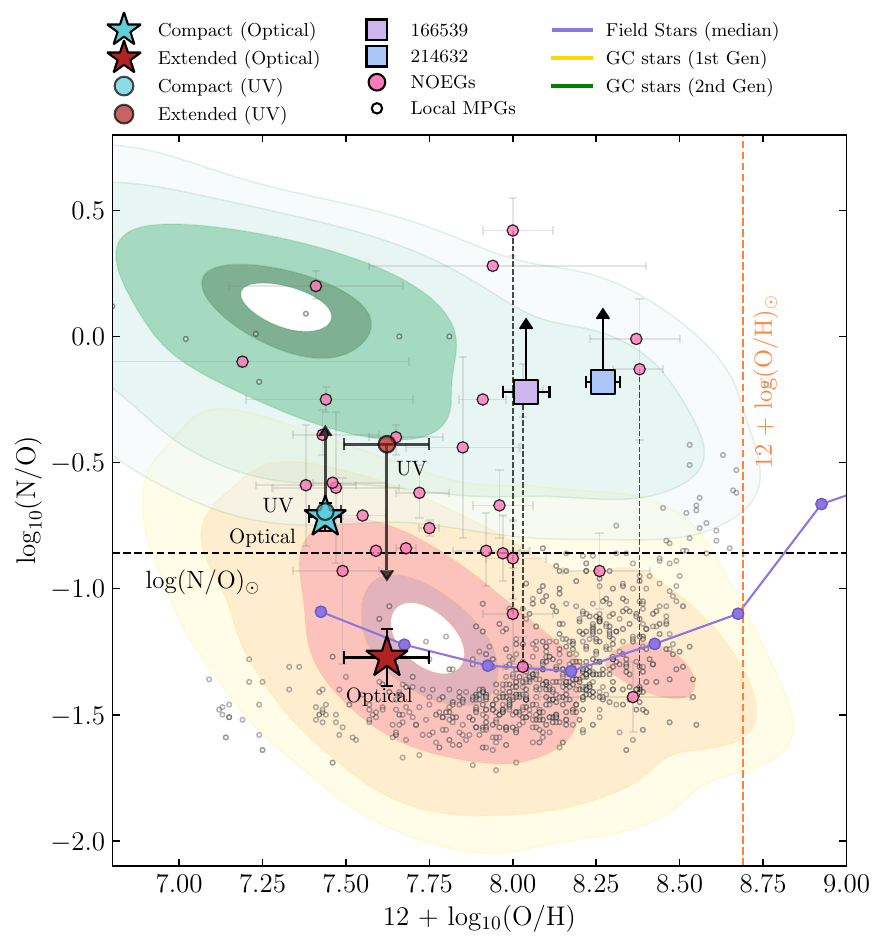}
    \caption{Distributions of the compact and extended morphology galaxies in the N/O versus O/H plane, comparing nitrogen abundances derived from both optical and UV emission lines. The blue star/circle symbols denote the compact population, while the red star/circle symbols represent the extended population. The star symbols denote log(N/O) derived from optical emission lines. The circle symbols denote log(N/O) derived from UV emission lines. The pink circles represent nitrogen-enhanced galaxies (NOEGs) compiled from the literature 
    \citep{castellano_jwst_2024, cameron_nitrogen_2023, cameron_nebular_2024, schaerer_discovery_2024, isobe_jwst_2023, larson_ceers_2023, topping_deep_2024, topping_aurora_2025, ji_ga-nifs_2024, ji_connecting_2025, ubler_ga-nifs_2023,pascale_nitrogen-enriched_2023, james_vlt_2009, arellano-cordova_classy_2025, napolitano_seven_2025, navarro-carrera_constraints_2024, stiavelli_what_2025, zhang_potential_2025}. NOEG systems which have two abundance values are connected by dashed lines. Their higher abundances are derived for high-density components $(n_{\rm e} \gtrsim 10^5  \ \rm cm^{-3})$ decomposed from the observed spectra, and their lower abundances derived for low-density components $(n_{\rm e} < 10^4  \ \rm cm^{-3})$ \citep{ji_connecting_2025}. For comparison, the purple curve shows the median abundance trend of Milky Way field stars. Local globular cluster stars are shown as yellow (1st generation) and green (2nd generation) kernel density contours from \citep{abdurrouf_seventeenth_2022}.  
    Five contour levels are plotted corresponding to 5, 16, 50, 84, and 95 percentiles of the distributions. Local metal-poor galaxies are shown as small white circles \citep{izotov_chemical_2006, pilyugin_abundance_2012, annibali_chemical_2019, grossi_southern_2025}. The dashed horizontal and vertical lines indicate the solar N/O ratio and solar oxygen abundance, respectively from \citet{asplund_chemical_2021}. Two nitrogen enhanced clumpy galaxies from our parent sample are highlighted as square points and are discussed in Section~\ref{clumpy_sources}. The compact population exhibits super-solar N/O values in both the optical and UV diagnostics, lying in the vicinity of the NOEGs and second-generation globular cluster stars. 
    In the optical measurements, the compact population is enhanced by approximately $\sim0.7$ dex in N/O relative to the extended population. }
    \label{fig:log(N/O)_median}
\end{figure}

Figure~\ref{fig:log(N/O)_median} demonstrates a clear distinction between the compact and extended populations. The compact median stack exhibits a well-constrained significantly higher optical $\rm log(N/O)_{optical}$ abundance than the extended stack by $\sim0.7$ dex. Similarly, the lower limit on the $\rm log(N/O)_{UV}$ abundance of the compact population is $\sim0.2$ dex higher than solar nitrogen abundance value. The $\rm log(N/O)_{UV}$ abundance of the extended sources is unconstrained and serves as an approximate upper limit. Overall, the compact galaxies exhibit super-solar log(N/O) abundance ratios, whereas the extended galaxies remain sub-solar.

The compact population occupies the same region of the N/O--O/H plane as previously identified NOEGs, particularly in the low metallicity regions. Previous studies have shown that compact sources show lower metallicities \citep{tacchella_jwst_2023}. In contrast, the extended population follows the sequence defined by the metal poor galaxies in $\rm 8.0\lesssim12+log(O/H)\lesssim8.5$ and is consistent with the median trend of the Milky Way thick-disk stars from APOGEE. These results suggest that nitrogen enhancement is not ubiquitous among high-redshift galaxies, but instead preferentially associated with the compact population.

Interestingly, the compact stack also overlaps with the abundance distribution of second-generation Milky Way globular clusters derived from APOGEE DR17. These second-generation local globular cluster stars are characterized to have peculiar abundances \citep{belokurov_nitrogen_2023, ji_connecting_2025, gratton_what_2019,  bastian_multiple_2018, charbonnel_n-enhancement_2023, pascale_nitrogen-enriched_2023}. They show enhanced nitrogen, sodium, and aluminum, together with depleted carbon, oxygen, and magnesium. This similarity between the compact galaxy abundances and those of chemically anomalous globular cluster stars suggests that the physical processes responsible for producing nitrogen-rich stellar populations locally may already be operating within compact star-forming regions at high redshift. The observed connection between compact morphology of high redshift galaxies and nitrogen enhancement is further supported by previous studies of NOEGs, many of which are themselves intrinsically compact compared to typical star-forming galaxies \citep{schaerer_discovery_2024, harikane_jwst_2025, topping_aurora_2025, ji_connecting_2025}. Hence, these results, strengthen the emerging picture that compactness and nitrogen enhancement are closely linked.

One possible explanation is that compact galaxies host significantly higher star formation and stellar mass surface densities, concentrating stellar feedback and chemical enrichment into smaller and denser regions. The electron densities we derive from the [S~\textsc{ii}]$\lambda\lambda6716,6731$ ratio in the stacked spectra are consistent with the compact and extended populations within their uncertainties. However, [SII] traces the low-ionisation gas and saturates at $n_{\rm e} \geq 10^{3}$--$10^{4} \, \rm cm^{-3}$, and stacking further averages over the population. Without resolved high-density tracers such as C~\textsc{iii}] or N~\textsc{iii}], we cannot determine whether compact galaxies preferentially host a larger fraction of dense, highly ionized gas.
As shown in Figure~\ref{fig:SFR_density}, the compact population exhibits systematically higher star formation rate surface density, $\Sigma_{\rm SFR}$, than the extended galaxies. We compute $\Sigma_{\rm SFR} = {\rm SFR}_{10\,{\rm Myr}}/(2\pi R_{\rm e}^2)$, using star formation rates from the JADES galaxy stellar population catalogue \citep{duan_jwst_2026} and effective radii from the JADES DR5 morphology catalogue \citep{carreira_jwst_2026}. Radial dust attenuation gradients could in principle bias the measured sizes, but these are expected to be small at these redshifts. Further, \citet{danhaive_beyond_2026} find consistent rest-frame FUV and NUV sizes for galaxies at $z\approx4$--6, indicating weak or spatially uniform dust attenuation. The enhanced $\Sigma_{\rm SFR}$ of the compact galaxies indicates that intense star formation occurs on much smaller physical scales, which may naturally produce localized regions where nitrogen enrichment can accumulate more efficiently.
 
Several physical mechanisms could contribute to the standard observed nitrogen enhancement by different channels. At low metallicity, nitrogen production may originate from rapidly rotating massive stars through rotational mixing and stellar winds \citep{matteucci_considerations_1986}. Intermediate mass stars may also contribute on relatively short timescales which can cause a large scatter in N/O values \citep{henry_cosmic_2000, meynet_stellar_2002, kobayashi_origin_2020}. At higher metallicities,  delayed nitrogen enrichment from winds of asymptotic giant branch (AGB) stars also becomes crucial \citep{henry_cosmic_2000, rai_unveiling_2026}. Alternatively, differential galactic winds may preferentially remove oxygen-rich material produced by CCSNe retaining nitrogen-enriched gas, thereby increasing the observed N/O ratio \citep{mcclymont_globular_2026}.

But these channels could vary for different star clusters, which could be present in the denser regions of compact sources. The compact galaxies may additionally host localized chemical anomalies. If super star clusters, proto-globular clusters, or nuclear star clusters dominate the star formation within these systems, nitrogen enrichment may become confined to dense central regions rather than being uniformly distributed throughout the galaxy \citep{james_vlt_2009, pascale_nitrogen-enriched_2023, ji_ga-nifs_2024, ji_connecting_2025}. Such chemical stratification naturally explains why nitrogen enhancement appears preferentially in compact galaxies and is largely absent in the more extended population.  

Overall, our results support a scenario in which nitrogen enhancement is fundamentally linked to compact, high-density star formation. Rather than representing a global property of early galaxies, the elevated N/O abundances may trace localized, chemically enriched regions associated with dense stellar clusters, providing a natural connection between compact galaxy morphology, intense star formation, and the emergence of chemically anomalous stellar populations in the early Universe.

\begin{figure}
	\includegraphics[width=1\columnwidth]{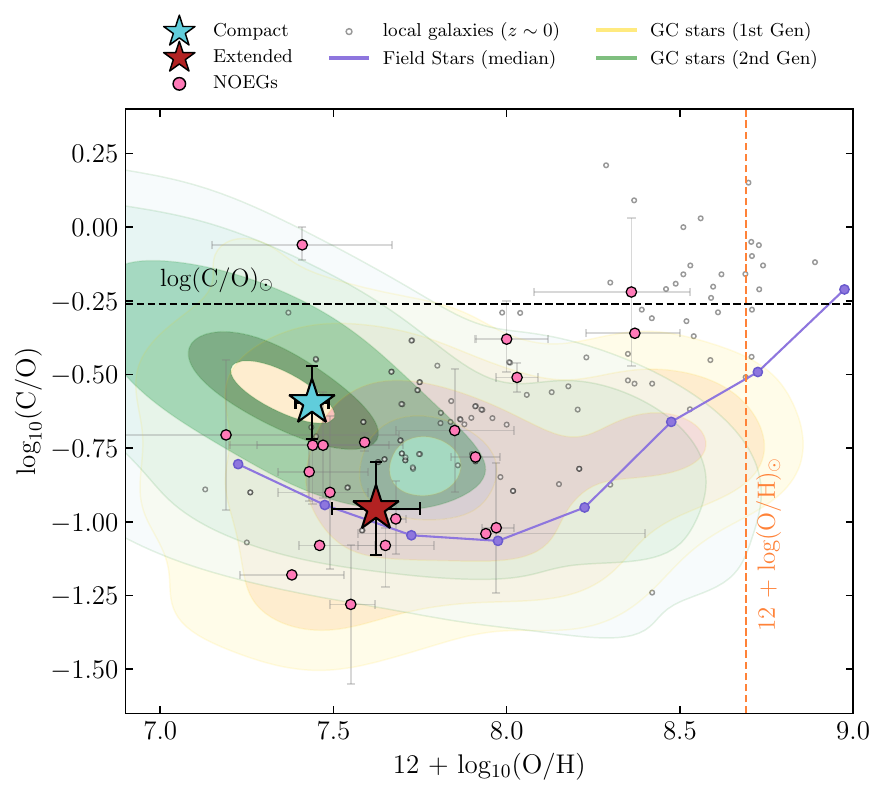}
    \caption{Distributions of the compact and extended morphology galaxies in the C/O versus O/H plane, comparing carbon abundances derived from both optical and UV emission lines. The blue star symbol denotes the compact population, while the red star symbol represents the extended population. which are derived from UV emission lines. The pink circles represent nitrogen-enhanced galaxies (NOEGs) compiled from the literature \citep{castellano_jwst_2024, cameron_nitrogen_2023, cameron_nebular_2024, schaerer_discovery_2024, isobe_jwst_2023, larson_ceers_2023, topping_deep_2024, topping_aurora_2025, ji_ga-nifs_2024, ji_connecting_2025, ubler_ga-nifs_2023,pascale_nitrogen-enriched_2023, james_vlt_2009, arellano-cordova_classy_2025, napolitano_seven_2025, navarro-carrera_constraints_2024, stiavelli_what_2025, zhang_potential_2025}. For comparison, the purple curve shows the median abundance trend of Milky Way field stars. Local globular cluster stars are shown as yellow (1st generation) and green (2nd generation) kernel density contours from \citep{abdurrouf_seventeenth_2022}.  The contours correspond to probability distributions computed using the \texttt{KDEPLOT} function from the \texttt{SEABORN} package. Five contour levels are plotted corresponding to 5, 16, 50, 84, and 95 percentiles of the distributions. Local metal-poor galaxies are shown as small white circles \citep{izotov_chemical_2006, pilyugin_abundance_2012, annibali_chemical_2019, grossi_southern_2025}. The dashed horizontal and vertical lines indicate the solar C/O ratio and solar oxygen abundance, respectively from \citet{asplund_chemical_2021}. Both the compact and extended populations exhibit sub-solar C/O values. The compact population follows the C/O–O/H trend defined by the literature NOEGs and local metal-poor galaxies. Furthermore, the compact population is enhanced by approximately $\sim0.53$ dex in C/O relative to the extended population. }
    \label{fig:log(C/O)_median}
\end{figure}

\subsection{Elevated C/O}

We detect the C~\textsc{iii}]$\lambda\lambda1907,1910$ emission line in the median R100 stacked spectra of both the compact and extended galaxy populations, enabling us to derive their $\rm C^{2+}/O^{2+}$ abundance ratios. We compare the resulting C/O abundances with measurements for NOEGs from the literature \citep{castellano_jwst_2024, cameron_nitrogen_2023, cameron_nebular_2024, schaerer_discovery_2024, isobe_jwst_2023, larson_ceers_2023, topping_deep_2024, topping_aurora_2025, ji_ga-nifs_2024, ji_connecting_2025, ubler_ga-nifs_2023,pascale_nitrogen-enriched_2023, james_vlt_2009, arellano-cordova_classy_2025, napolitano_seven_2025, navarro-carrera_constraints_2024, stiavelli_what_2025, zhang_potential_2025}, Milky Way globular clusters and field stars from APOGEE \citep{abdurrouf_seventeenth_2022} and nearby metal-poor galaxies \citep{izotov_chemical_2006, pilyugin_abundance_2012, annibali_chemical_2019, grossi_southern_2025}.

Figure~\ref{fig:log(C/O)_median} shows that both the compact and extended populations exhibit sub-solar C/O ratios. However, compact galaxies display significantly elevated carbon abundances, with log(C/O) enhanced by approximately $0.5$ dex relative to the extended population. Despite this enhancement, the compact stacks remain below the solar C/O ratio, consistent with the majority of NOEGs reported in the literature.

The compact population occupies the same region of the C/O--O/H plane as the known NOEGs and local ($z\sim0$) metal-poor galaxies. Furthermore, it follows the observed increase of C/O with metallicity, a trend seen across both nearby galaxies and stellar populations. In contrast, the extended population lies systematically below this sequence and does not follow the observed C/O--O/H relation, indicating comparatively weaker carbon enrichment.

The observed enhancement in the compact population suggests a larger contribution from delayed carbon production relative to oxygen. Unlike oxygen, which is produced predominantly by massive stars and released in short timescales, a significant fraction of carbon originates from secondary nucleosynthesis channels that enrich the interstellar medium on longer timescales \citep{garnett_evolution_1995, kobayashi_origin_2020}. The increasing C/O ratio with metallicity observed in Milky Way stars \citep{ji_connecting_2025} is consistent with this delayed carbon enrichment, and the compact galaxies follow this same evolutionary sequence.

The sub-solar C/O ratios also imply enhanced N/C abundances, consistent with the strong enrichment observed in the compact population. Such abundance patterns can naturally arise from CNO processing in asymptotic giant branch (AGB) stars, where the degree of carbon conversion into nitrogen depends on stellar mass and metallicity, producing a broad range of N/C ratios \citep[e.g.,][]{herwig_evolution_2004, karakas_stellar_2007, ventura_yields_2013}. Alternatively, the enhanced N/C could originate from other enrichment channels proposed for NOEGs, including Wolf--Rayet (WR) stars, tidal disruption events, or supermassive stars \citep{cameron_nitrogen_2023, isobe_empress_2023, watanabe_empress_2024}. The similarity between the compact galaxies and previously identified NOEGs therefore suggests that they may have experienced comparable chemical enrichment histories.

\subsection{Star Formation Rate Surface Density - $\Sigma_{\rm SFR}$}

We compare the star formation rate surface densities of the two morphological populations to investigate whether the compact and extended systems occupy distinct star-forming regimes. We calculate the star formation rate surface density as
$\Sigma_{\rm SFR} = {\rm SFR_{10\  Myr}}/2\pi R_{\rm e}^2$, where $\rm SFR_{10\  Myr}$ is the star formation rate averaged over the past 10 Myr and $R_{\rm e}$ is the effective radius. We obtain $\Sigma_{\rm SFR, \ compact} = 8.56 \pm 1.41  \ M_\odot \rm yr^{-1} kpc^{-2}$ and $\Sigma_{\rm SFR, \ extended} = 0.06 \pm 0.01 \ M_\odot \rm yr^{-1} kpc^{-2}$. Thus, the compact population has a higher star formation rate surface density than the extended population based on our sample selections, indicating that its recent star formation is concentrated within a much smaller physical area.

We additionally quantify the recent burstiness of the star formation using the ratio
$\rm SFR_{10 \rm \ Myr}/\rm SFR_{100\ \rm Myr}$, where the subscripts denote the timescale over which the star formation rate is averaged. We find $\rm SFR_{10}/\rm SFR_{100, \rm \ compact} = 2.13 \pm 0.21$ for the compact population and $\rm SFR_{10}/\rm SFR_{100, \rm \ extended} = 1.68 \pm 0.11$ for the extended population. The larger ratio for the compact systems suggests that, in addition to having more spatially concentrated star formation, they have experienced a stronger recent increase in their star formation activity relative to their longer-term average. Together, the higher $\Sigma_{\rm SFR}$ and elevated burstiness point towards a picture in which the compact systems are undergoing particularly intense and spatially concentrated recent star formation.
Further analysis of the relationships between galaxy properties and morphology are presented in Carreira et al. (in prep.).

\begin{figure}
	\includegraphics[width=1\columnwidth]{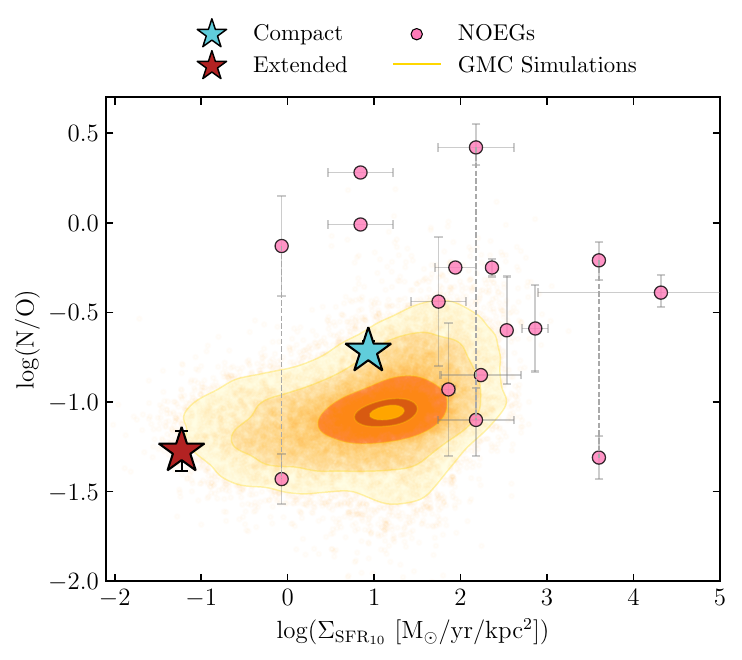}
    \caption{The nitrogen-to-oxygen abundance ratio as a function of the SFR surface density, $\Sigma_{\rm SFR}$, for the compact and extended stacked populations. The blue and red stars represent the compact and extended populations, respectively, while the pink circles show nitrogen-enhanced galaxies (NOEGs) compiled from the literature \citep{castellano_jwst_2024, cameron_nitrogen_2023, cameron_nebular_2024, schaerer_discovery_2024, isobe_jwst_2023, larson_ceers_2023, topping_deep_2024, topping_aurora_2025, ji_ga-nifs_2024, ji_connecting_2025, ubler_ga-nifs_2023, pascale_nitrogen-enriched_2023, james_vlt_2009, arellano-cordova_classy_2025, napolitano_seven_2025, navarro-carrera_constraints_2024, stiavelli_what_2025, zhang_potential_2025}. NOEG systems with two abundance measurements are connected by dashed lines. In these systems, the higher N/O abundances are derived from high-density components ($n_{\rm e} \gtrsim 10^5~\rm cm^{-3}$) decomposed from the observed spectra, while the lower abundances are derived from low-density components ($n_{\rm e} < 10^4~\rm cm^{-3}$) \citep{ji_connecting_2025}. For comparison, we show the simulated giant molecular clouds (GMCs) from \texttt{THESAN-ZOOM} simulations from \citet{mcclymont_span_2026}. The yellow and orange contours indicate the 5th, 16th, 50th, 84th, and 95th percentiles of the GMC distributions, with the individual GMCs shown as orange circles. The nitrogen-rich GMCs arise at the onset of a fresh starburst, when pre-polluted gas undergoes gravitational collapse, and occupy $\Sigma_{\rm SFR}$ values comparable to those of the observed NOEGs. The compact stack lies closer to the locus occupied by these nitrogen-rich GMCs and the NOEGs in the literature, suggesting a closer correspondence between the compact population and the conditions associated with nitrogen enrichment.}

    \label{fig:SFR_density}
\end{figure}

The difference in $\Sigma_{\rm SFR}$ between the two populations provides an important physical distinction when considering the origin of the unusual abundance patterns observed in the nitrogen-enhanced galaxies. The compact systems have $\Sigma_{\rm SFR}$ more than two orders of magnitude higher than the extended population, placing their star formation in a substantially denser regime. Such concentrated star formation can create environments in which stellar ejections and chemically enriched material are deposited into a relatively small volume, potentially allowing localized abundance enhancements to become prominent in UV emission. In particular, the combination of high $\Sigma_{\rm SFR}$, recent burstiness, and compact morphology provides conditions that may favor the formation or observational dominance of chemically enriched star-forming regions.

This distinction is also illustrated in Figure~\ref{fig:SFR_density}, where we compare the $\Sigma_{\rm SFR}$ values of our compact and extended stacks with measurements for nitrogen-enhanced galaxies (NOEGs) in the literature \citep{ji_connecting_2025}. The compact population occupies a $\Sigma_{\rm SFR}$ regime comparable to that of the observed NOEGs, whereas the extended population lies substantially below this regime. This similarity suggests that the extreme UV emission-line properties of NOEGs may preferentially arise in systems where recent star formation is sufficiently concentrated to produce high local star formation intensities.

We also compare the nitrogen-rich giant molecular clouds (GMCs) identified in the \texttt{THESAN-ZOOM} simulations of \citet{mcclymont_span_2026} with the star formation rate surface densities of the compact and extended stacks. As shown in Figure~\ref{fig:SFR_density}, these nitrogen-rich GMCs occupy $\Sigma_{\rm SFR}$ values comparable to those observed in NOEGs and to those of our compact population. In the simulations, the nitrogen-rich GMCs form at the onset of a new starburst, when previously enriched gas undergoes gravitational collapse. This gas has been polluted by AGB stars during an earlier episode star-burst, such that the newly collapsing molecular clouds inherit enhanced nitrogen abundances from the preceding stellar population. The resulting nitrogen-rich GMCs can subsequently dominate the UV line emission, producing the extreme abundance patterns characteristic of NOEGs \citep{mcclymont_span_2026, rai_unveiling_2026}.

The simulated scenario provides a useful physical framework for interpreting the difference between our compact and extended populations. In this picture, the high $\Sigma_{\rm SFR}$ of compact systems does not necessarily imply that the entire galaxy is uniformly nitrogen enriched. Rather, the concentrated star formation may increase the contribution of localized, chemically enriched regions to the integrated spectrum. If these regions contain gas that was enriched by intermediate-mass AGB stars during an earlier episode of star formation, a subsequent burst could illuminate this pre-enriched material through strong UV emission lines. The enhanced nitrogen abundance would therefore reflect the chemical history of the gas participating in the current burst, rather than simply the overall metallicity of the galaxy.


This interpretation is also consistent with the decoupling between the nitrogen and oxygen abundances of the compact population. Secondary nitrogen production by intermediate-mass AGB stars occurs on delayed timescales following the formation of their progenitors, allowing nitrogen-rich material to be released after an earlier burst of star formation. At the same time, continued inflow of relatively pristine gas can dilute the overall gas-phase metallicity and lower the oxygen abundance. Consequently, a system can exhibit elevated N/O while maintaining relatively low O/H. Alternatively, nitrogen enrichment may be driven by the winds of massive stars, such as Wolf--Rayet (WR) stars or supermassive stars, in which case the N/O-enhanced phase precedes the onset of core-collapse supernovae. The high $\Sigma_{\rm SFR}$ of the compact population is also consistent with these scenarios. Models of supermassive star formation typically require very high gas densities to sustain the high mass accretion rates needed for their growth, whether through gas accretion \citep{gieles_concurrent_2018} or runaway stellar collisions \citep{rantala_frost-clusters_2024}. Similarly, in the WR scenario, \citet{fukushima_impacts_2024} find that massive, dense star clusters are favourable for WR stars to dominate the enrichment. However, we note, that the high $\Sigma_{\rm SFR}$ is not itself evidence that compactness directly causes nitrogen enhancement; rather, it may trace the dense, rapidly star-forming environments in which chemically enriched gas can contribute disproportionately to the observed emission.


Taken together, the higher $\Sigma_{\rm SFR}$, stronger recent star formation, and similarity to the star-forming conditions of nitrogen-rich GMCs in \texttt{THESAN-ZOOM} suggest that the compact population may preferentially host the dense and rapidly evolving star-forming regions required to produce the extreme abundance patterns observed in NOEGs \citep{mcclymont_span_2026}. The combination of spatial concentration and delayed chemical enrichment offers a plausible explanation for why nitrogen enhancement is associated with the compact population, while also naturally allowing the nitrogen-rich systems to retain relatively low global oxygen abundances.

\begin{figure*}
    \centering
	\includegraphics[width=1\linewidth]{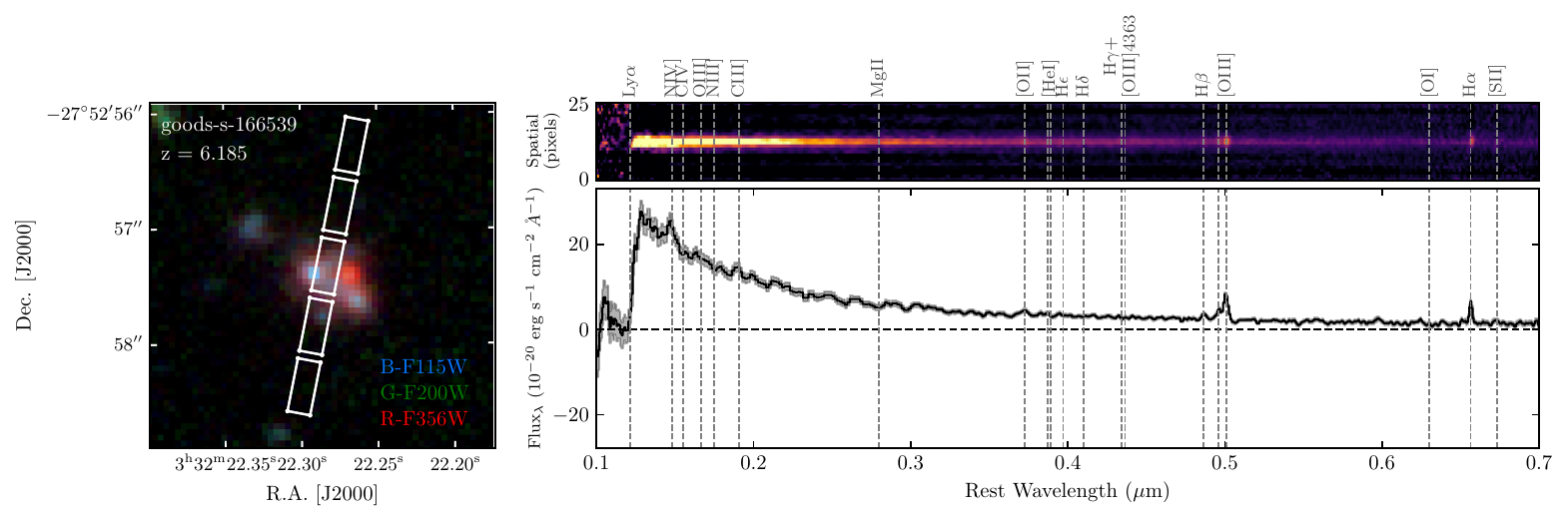}
    \includegraphics[width=1\linewidth]{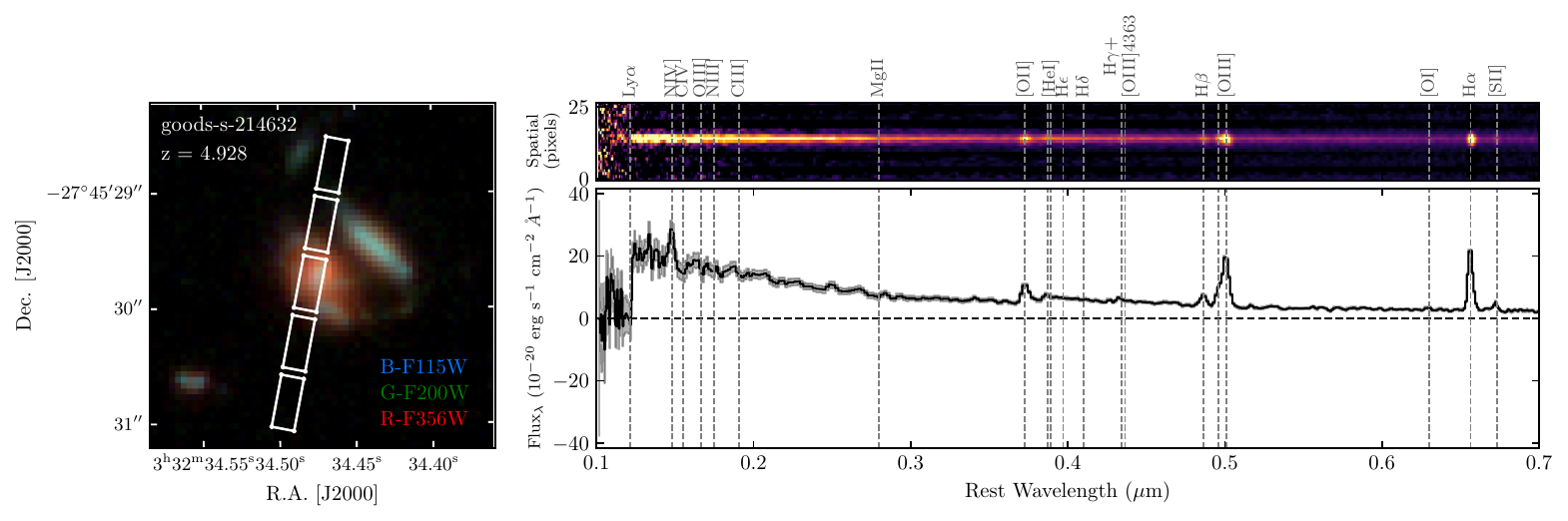}
    \caption{NIRCam images and NIRSpec spectra of two clumpy nitrogen loud galaxies in JADES DR4: \textit{goods-s-mediumjwst-166539}, \textit{goods-s-mediumjwst-214632}. \textbf{Left Panel:} RGB NIRCam image with F115W, F200W and F356W filters and MSA slits overplotted. \textbf{Right Panel:} 1D and 2D NIRSpec low resolution prism spectra with key emission lines. Both these sources appear clumpy in nature and have a strong N~\textsc{iv}]$\lambda\lambda1483,1486$ detections.}
    \label{fig:clumpy_spectra}
\end{figure*}
\subsection{Two N-loud Galaxies}\label{clumpy_sources}

We find two galaxies which are N-loud. They have visually high N~\textsc{iv}]$\lambda\lambda1483,1486$ line and no other detected UV lines. These are IDs: \textit{goods-s-mediumjwst\_166539} and \textit{goods-s-mediumjwst\_214632}. One of these sources, MSA ID--166539, has been discussed in \citet{morel_discovery_2025}. We have R100 for both these sources but R1000 only for \textit{goods-s-mediumjwst\_166539}. For the sake of consistency, we derive all fluxes from the R100 observations.
We detect the N~\textsc{iv}]$\lambda\lambda1483,1486$ line with a $S/N=4.5$ for the \textit{goods-s-mediumjwst\_166539} source and a $S/N=4.0$ for the \textit{goods-s-mediumjwst\_214632} source. We do not detect any other UV emission lines and obtain $3\sigma$ upper limits for O~\textsc{iii}]$\lambda\lambda1661,1666$,  N~\textsc{iii}]$\lambda\lambda1747$-$1754$, and C~\textsc{iii}]$\lambda\lambda1907,1909$. We also detect the high ionisation lines in the optical, including H$\alpha$, H$\beta$, [O~\textsc{ii}]$\lambda\lambda3726,3729$, 
[O~\textsc{iii}]$\lambda\lambda5007,4959$, and [S~\textsc{ii}]$\lambda\lambda6716,6731$. We convert the line fluxes to integrated fluxes and obtain their errors through Monte Carlo simulations. We then calculate the Balmer decrement which gives us $\rm H\alpha / H\beta = 3.05$ for \textit{goods-s-mediumjwst\_166539} and $\rm H\alpha / H\beta = 4.72$ for \textit{goods-s-mediumjwst\_214632} and use this to dust-correct the integrated fluxes.

Medium-resolution (R1000) spectra are available only for \textit{goods-s-mediumjwst\_166539}, but we do not detect the auroral [O~\textsc{iii}]$\lambda4363$ line, and therefore cannot derive a direct electron temperature. We thus adopt a constant $T_{\rm e}=10^4$~K and assume the low-density regime with $n_{\rm e}=100$~cm$^{-3}$ for both sources.

Drawing metallicity measurements from \citet{isobe_jades_2026}, which are based on the strong-line method and the DR4 flux catalogue, we report that \textit{goods-s-mediumjwst\_166539} has $\rm 12+log(O/H)=8.04\pm0.07$ and \textit{goods-s-mediumjwst\_214632} has $\rm 12+log(O/H)=8.27\pm0.05$.

We then calculate the nitrogen abundance $\rm N^{3+}/O^{2+}$ using N~\textsc{iv}]~$\lambda\lambda1483,1486$/O~\textsc{iii}]~$\lambda\lambda1661,1666$ line ratio according to 
\begin{equation}\label{eq:N3+/O2+_clumpy} 
    {\rm N^{3+} \over {O^{2+}}} = {I_{\lambda\lambda1486,1483} \over I_{\lambda\lambda1661,1666}} \times {\epsilon_{\lambda\lambda1661,1666} \over \epsilon_{\lambda\lambda1486,1483}}
\end{equation}
where the $I_\lambda$ denotes the integrated emission-line fluxes measured from the Monte Carlo \texttt{pPXF} and the $\epsilon_\lambda$ represent the corresponding emissivities calculated from \texttt{PyNeb}. We do not detect the O~\textsc{iii}]~$\lambda\lambda1661,1666$ line in both sources so we use the $3\sigma$ upper limits for this line. This then gives us the lower limit for the nitrogen abundances that we calculate as

\begin{equation}\label{eq:N/O_clumpy} 
    \rm log\left({\rm N \over {O}}\right)> \rm log\left(\rm {N^{3+} \over O^{2+}} \right) 
\end{equation}

Using this approach, we derive $\rm log(N/O) > -0.22$ for \textit{goods-s-mediumjwst\_166539} and $\rm log(N/O)  >-0.18$ for \textit{goods-s-mediumjwst\_214632}. We see that both these sources have a really high nitrogen abundance comparable to the other N-enhanced galaxies in the literature as shown in Figure~\ref{fig:log(N/O)_median} \citep{castellano_jwst_2024, cameron_nitrogen_2023, cameron_nebular_2024, schaerer_discovery_2024, isobe_jwst_2023, larson_ceers_2023, topping_deep_2024, topping_aurora_2025, ji_ga-nifs_2024, ji_connecting_2025, ubler_ga-nifs_2023,pascale_nitrogen-enriched_2023, james_vlt_2009, arellano-cordova_classy_2025, napolitano_seven_2025, navarro-carrera_constraints_2024, stiavelli_what_2025, zhang_potential_2025}

Further, \textit{goods-s-mediumjwst\_166539} is classified as an extended source in our sample, while \textit{goods-s-mediumjwst\_214632} was excluded from the morphology-based samples because it lacks an R1000 stack. However, the NIRCam images of both galaxies reveal a clumpy morphology, consisting of several distinct compact clumps \citep{puskas_clumpy_2026}. While their overall light distribution is spatially extended, consistent with the classification of \textit{goods-s-mediumjwst\_166539}, the individual clumps are compact. Since the effective radii are derived from single-component S\'ersic fits with \texttt{pysersic} \citep{carreira_jwst_2026}, they trace the global extent of the light distribution rather than the sizes of the individual clumps. Nitrogen enhancement in these galaxies may therefore originate in compact, dense star-forming clumps embedded within an extended system, in line with the localized enrichment scenario discussed above. Spatially resolved spectroscopy (e.g. with the NIRSpec IFU) is needed to establish whether the nitrogen enrichment is associated with individual clumps.


\section{Conclusions}\label{sec:conclusions}

In this study, we investigate the correlation between chemical abundances of high redshift galaxies from JADES DR4 R100 low-resolution prism and R1000 medium-resolution grating spectra and their morphological properties via median stacking \citep{curtis-lake_jades_2026, eisenstein_jades_2025, scholtz_jades_2025, deugenio_jades_2024, carreira_jwst_2026, duan_jwst_2026, alberts_jwst_2026, hainline_jwst_2026, johnson_jwst_2026, robertson_jwst_2026}. We divide our parent spectroscopic sample of 657 $4<z<7$ galaxies into compact, intermediate and extended sources based on their stellar mass-normalized rest-frame UV effective radii. We model emission lines at rest-frame UV and optical wavelengths, focusing on nebular emission lines of carbon, nitrogen, and oxygen. We compare the gas phase metallicities, nitrogen-to-oxygen and carbon-to-oxygen ratios between the compact and extended stacked spectra. We further compare the derived chemical abundances to those in the simulated giant molecular clouds (GMCs) from the THESAN-ZOOM simulations.

Our main results are summarized as follows:
\begin{enumerate}

    \item \textbf{Gas-phase metallicities:} We derive gas-phase metallicities for the compact and extended R1000 stacks using the direct $T_{\rm e}$ method, with electron temperatures determined from the auroral and strong oxygen lines. We find $12+\log({\rm O/H})_{\rm compact}=7.44\pm0.05$ ($0.06Z_{\odot}$) and $12+\log({\rm O/H})_{\rm extended}=7.62\pm0.13$ ($0.09Z_{\odot}$), both sub-solar, adopting $12+\log({\rm O/H})_\odot=8.69$ \citep{asplund_chemical_2021}. This suggests that the compact galaxies host more metal-poor ionized gas, whereas the extended population has undergone comparatively greater chemical enrichment.

    \item \textbf{Nitrogen enrichment:} We derive nitrogen-to-oxygen abundances from the R1000 spectra using the optical [N~\textsc{ii}]$\lambda\lambda6548,6584$ and [O~\textsc{ii}]$\lambda\lambda3726,3729$ lines, finding $\log({\rm N/O})_{\rm compact,\,optical}=-0.72\pm0.05$ and $\log({\rm N/O})_{\rm extended,\,optical}=-1.27\pm0.11$. We also derive N/O from the R100 spectra using the rest-frame UV N~\textsc{iv}]$\lambda\lambda1483,1486$, N~\textsc{iii}]$\lambda\lambda1747$--$1754$, and O~\textsc{iii}]$\lambda\lambda1661,1666$ lines, obtaining $\log({\rm N/O})_{\rm compact,\,UV}>-0.69$ and $\log({\rm N/O})_{\rm extended,\,UV}<-0.43$. With both the optical and UV tracers, the compact stack shows a super solar nitrogen abundance with substantially higher nitrogen abundance than the extended stack in the optical tracer. The $\log({\rm N/O})_{\rm extended,\,UV}$ value remains unconstrained. Combined with its low metallicity, the super-solar N/O of the compact stack \citep[$\log({\rm N/O})_\odot=-0.86$;][]{asplund_chemical_2009} places it in the same region of the N/O--O/H plane as previously identified nitrogen-enhanced galaxies (NOEGs).
    
    \item \textbf{Carbon Enrichment:} We derive the carbon abundances from the UV rest-frame R100 spectra using the C~\textsc{iii}]$\lambda\lambda1907,1909$ and O~\textsc{iii}]$\lambda\lambda1661,1666$ lines as $\rm log(C/O)_{compact, \ UV} = -0.58\pm0.11$ for the compact stack and $\rm log(C/O)_{extended, \ UV} = -1.02\pm0.14$ for the extended stack. We find that the compact stacks show significantly higher nitrogen and carbon abundances relative to oxygen than the extended stacks. The nitrogen abundances of the compact stacks have both sub-solar metallicities and log(C/O) and occupy the same region of the C/O--O/H plane as the previously identified NOEGs. The carbon enhancement could suggest a contribution from delayed carbon production relative to the oxygen originating from secondary nucleosynthesis channels that enrich the ISM following a similar enrichment sequence compared to MW stars.

    \item \textbf{Comparison of Star Formation Rate Surface Density:} The compact stack show significantly higher $\Sigma_{\rm SFR}$ ($\Sigma_{\rm SFR, \ compact}=8.56\pm1.41$, $\Sigma_{\rm SFR, \ extended}=0.06\pm0.01$), indicating more intense and spatially concentrated star formation. The resulting high stellar-feedback rates may promote localized enrichment within the dense star-formation regions of these systems. The enhanced nitrogen abundances may arise from the delayed release of secondary nitrogen by intermediate-mass AGB stars following an earlier episode of star formation, while continued inflow of relatively pristine gas can dilute the overall gas-phase metallicity. This combination provides a possible explanation for the elevated N/O ratios observed in the compact population despite its lower overall oxygen abundance, suggesting that nitrogen enrichment and global metal enrichment need not proceed simultaneously.
    
    \item \textbf{Comparison with Simulations:} We compare our chemical abundance values to the simulated abundances in the nitrogen-rich GMCs from the THESAN-ZOOM simulations from \citet{mcclymont_span_2026}. We find that the compact stack occupies a $\Sigma\rm SFR$ comparable to the nitrogen-rich GMCs and the NOEGs. The nitrogen enhancement could arise within localized, chemically enriched regions associated with dense star-forming clusters, with the elevated SFR and $\Sigma\rm SFR$ of the compact stack providing conditions conducive to such localized enrichment. We further see that the compact stacks show higher burstiness compared to the extended stacks. Here, we define burstiness as the ratio of the recent ($\sim10$ Myr) star-formation rate to the longer-timescale ($\sim100$ Myr) average SFR, such that higher values indicate a more rapidly varying or burst-dominated star-formation history. This could suggest that the bursty nature of the galaxies could have a period of delayed nitrogen enrichment from AGB stars over short timescales. These star formation bursts would also maintain the feedback loop of pristine gas inflow from CCSNe and nitrogen enrichment from AGB stars.

    \item \textbf{Two N-loud Galaxies:} We find two galaxies in our parent sample which show strong N~\textsc{iv}$\lambda\lambda1483,1486$. Further, when looking chemical abundances of the two nitrogen loud galaxies \textit{goods-s-mediumjwst\_166539} and \textit{goods-s-mediumjwst\_214632}, we see that these galaxies show super solar nitrogen (log(N/O)) abundances ($\rm log(N/O)_{166539} >-0.22$, $\rm log(N/O)_{214632} >-0.18$). But looking at the morphology of these sources, they appear clumpy in nature with multiple components spatially resolved. This shows us that studying the morphology is essential to understand the location of this nitrogen enrichment. Further investigation of such sources in spatially resolved spectra (IFU) with higher resolution would help to localize the N-enrichment seen in these clumpy galaxies.

\end{enumerate}

Our findings suggest that compact galaxies at high redshift overall show enhanced nitrogen and carbon abundances at lower metallicities compared to extended galaxies. The compact sources also show higher SFR, $\Sigma_{\rm SFR}$, and burstiness compared to the extended sources.

The higher burstiness of the compact sources suggests that they may undergo more rapid episodes of star formation. Following an initial burst, feedback from massive stars and CCSNe can drive the redistribution or expulsion of metal-enriched gas, while subsequent accretion of relatively pristine gas can dilute the ISM and lower the overall oxygen abundance. On longer timescales, delayed nitrogen production from intermediate-mass AGB stars can then enhance N/O without a corresponding increase in O/H. This combination of bursty star formation, gas dilution, and delayed nitrogen enrichment could therefore produce the elevated N/O ratios observed at relatively low metallicities in the dense, compact systems, consistent with the enrichment pathways proposed in \citet{rai_unveiling_2026}. Our results therefore suggest that the observed nitrogen enhancement may arise from the interplay between compact morphology, intense and bursty star formation, and the different enrichment timescales of oxygen and nitrogen.

Overall, our results support a scenario in which nitrogen enhancement is closely associated with compact, high-density star formation. The elevated N/O abundances may trace localized, chemically enriched regions associated with dense stellar clusters, linking compact galaxy morphology, intense and bursty star formation, and the emergence of chemically anomalous stellar populations in the early Universe. Future spatially resolved spectroscopy of larger samples will be essential for establishing whether the enhanced N/O ratios are indeed confined to the most compact star-forming regions and for determining the spatial relationship between nitrogen enrichment and stellar structure. Deeper observations targeting additional nitrogen and auroral lines will further constrain the ionisation conditions, electron temperatures, and abundance patterns, helping to distinguish between delayed enrichment from intermediate-mass stars, stellar feedback, and gas inflows as the dominant mechanisms. Expanding this analysis to larger samples spanning a broader range of masses, sizes, and redshifts will also establish whether the compactness–nitrogen enhancement connection represents a common phase of early galaxy evolution or a distinct population of chemically anomalous systems.

\section*{Acknowledgements}

UT acknowledges support from a PhD studentship awarded by the University of Cambridge, UK Science and Technology Facilities Council (STFC). XJ, YI, RM, FDE acknowledge ERC Advanced Grant 695671 ``QUENCH'' and support by the Science and Technology Facilities Council (STFC) and by the UKRI Frontier Research grant RISEandFALL. RM acknowledges funding from a research professorship from the Royal Society. CC acknowledges support from the JWST/NIRCam Science Team contract to the University of Arizona, NAS5-02105, and JWST Programs 3215 and 5015. BER acknowledges support from the NIRCam Science Team contract to the University of Arizona, NAS5-02105, and JWST Program 3215. AJB acknowledges funding from the “FirstGalaxies” Advanced Grant from the European Research Council (ERC) under the European Union’s Horizon 2020 research and innovation program (Grant agreement No. 789056). QD acknowledges support from a PhD studentship awarded by Trinity College, University of Cambridge. MK acknowledges support from a PhD studentship awarded by the University of Cambridge Harding Distinguished Postgraduate Scholars Programme, UK Science and Technology Facilities Council (STFC) Center for Doctoral Training (CDT) in Data Intensive Science, and Girton College Cambridge. PR acknowledges support from the University of Texas at Austin Cosmic Frontier Center. J.S. acknowledges support by the Science and Technology Facilities Council (STFC), ERC Advanced Grant 695671 "QUENCH".
CS acknowledges support from the Science and Technology Facilities Council (STFC), by the ERC through Advanced Grant 695671 “QUENCH”, by the UKRI Frontier Research grant RISEandFALL. ST acknowledges support by the Royal Society Research Grant G125142. H\"U acknowledges support by the Max Planck Society through the Lise Meitner Excellence Program. H\"U acknowledges funding by the European Union (ERC APEX, 101164796). Views and opinions expressed are however those of the authors only and do not necessarily reflect those of the European Union or the European Research Council Executive Agency. Neither the European Union nor the granting authority can be held responsible for them.

\section*{Data Availability}

The inclusion of a Data Availability Statement is a requirement for articles published in MNRAS. Data Availability Statements provide a standardized format for readers to understand the availability of data underlying the research results described in the article. The statement may refer to original data generated in the course of the study or to third-party data analysed in the article. The statement should describe and provide means of access, where possible, by linking to the data or providing the required accession numbers for the relevant databases or DOIs.



\bibliographystyle{mnras}
\bibliography{references_new} 

\bsp	
\label{lastpage}
\end{document}